\documentclass[11pt,aps,pra,onecolumn]{revtex4-2}
\usepackage[english]{babel}
\usepackage{amsmath,amssymb,amsthm,amsfonts,bm,braket,physics,graphicx}
\usepackage{tikz}

\usepackage{xcolor}

\newcommand{\inner}[2]{\langle #1 | #2 \rangle}

\usepackage{hyperref}
\begin{document}

\title{
Lecture Notes on Information Scrambling, Quantum Chaos, and Haar-Random States
}

\author{Marcin P{\l}odzie\'n$^{1}$}
\email{marcin.plodzien@qilimanjaro.tech}
\affiliation{$^1$
  Qilimanjaro Quantum Tech, Carrer de Veneçuela 74, 08019
Barcelona, Spain
  }

\date{\today}

\begin{abstract}
Information scrambling—the process by which quantum information spreads and
becomes effectively inaccessible—is central to modern quantum statistical
physics and quantum chaos.  These lecture notes provide an introduction to
information scrambling from both static and dynamical perspectives.  The
spectral properties of reduced density matrices arising from Haar-random states
are developed through the geometry of the unitary group and the universal
results of random matrix theory.  This geometric framework yields universal,
model-independent predictions for entanglement and spectral statistics,
capturing generic features of quantum chaos without reference to microscopic
details.  Dynamical diagnostics such as the spectral form factor and
out-of-time-ordered correlators further reveal the onset of chaos in
time-dependent evolution.

The notes are aimed at advanced undergraduate and graduate students in physics,
mathematics, and computer science who are interested in the connections between
quantum chaos, information dynamics, and quantum computing.  Concepts such as
entanglement growth, Haar randomness, random-matrix statistics, and unitary
$t$-designs are introduced through their realization in random quantum
circuits and circuit complexity.  The same mathematical framework also underpins
modern quantum-device benchmarking, where approximate unitary designs and random
circuits translate geometric ideas into quantitative tools for assessing
fidelity, noise, and scrambling efficiency in real quantum processors.
\end{abstract}

\maketitle

{
\section{Introduction}

The study of quantum information scrambling lies at the intersection of quantum chaos,
thermalization, and quantum information theory~\cite{DAlessio2016,Deutsch2018ETHreview,Liu2018}.
Scrambling refers to the rapid delocalization of initially localized information across many
degrees of freedom, making it inaccessible to any small subsystem.  
Although the global evolution is unitary, local observables often relax to
thermal values, reflecting a form of effective irreversibility familiar from
statistical mechanics
\cite{Swingle2016MeasuringScrambling,Swingle2018UnscramblingOTOC,XuSwingle2024ScramblingOTOC,Swingle2018Boulder}.

A particularly clean mathematical model of a fully scrambled state is provided
by the \emph{Haar ensemble}---the uniform distribution over pure states in a
Hilbert space~\cite{Mele2024}. 
No physical system can literally explore this ensemble, but many-body systems
with chaotic dynamics behave, after sufficiently long times, as if their local
subsystems had been drawn from it. Haar-random unitaries capture the long-time limit of chaotic dynamics:
chaotic Hamiltonians, random quantum circuits, and even black holes all appear to
approach this limit in their late-time entanglement structure
\cite{hayden2007black,hosur2016chaos,nahum2017quantum,ho2018chaos,Sekino2008,maldacena2016bound}.  In this setting, the key objects are not traditional observables but the
eigenvalues of \emph{reduced density matrices}, which encode entanglement and
effective thermality.  For Haar-random states, these eigenvalues obey universal
random-matrix statistics
\cite{mehta2004random,Forrester1993,Tao2012TopicsRMT}, leading to
Page’s theorem~\cite{page1993average} and providing the statistical foundation
for quantum typicality and the eigenstate thermalization hypothesis (ETH)
\cite{deutsch1991quantum,srednicki1994chaos,DAlessioRigol2016,Deutsch2018ETHreview}.

These notes introduce the reader to the mathematical tools and physical intuition needed to work
with these ideas.  
We begin by reviewing the unitary group \(U(D)\) as the natural configuration
space for closed quantum evolution, emphasizing its geometry as a compact Lie
group with a unique uniform measure: the Haar measure.  
When a Haar-random state on a bipartite system is partially traced, the reduced
state belongs to the fixed-trace Wishart ensemble.  
We derive the associated joint eigenvalue distribution, discuss eigenvalue
repulsion using the Coulomb-gas analogy, and show how the Marchenko--Pastur law
emerges in the large-dimension limit.  
These tools are then used to compute spectral moments, purity, and entanglement
entropies, culminating in Page’s theorem and a quantitative formulation of
typicality and local thermality. A recurring theme is the role of concentration-of-measure phenomena in large
Hilbert spaces, which help explain why so many distinct chaotic systems exhibit
nearly identical entanglement behavior.

Next, we focus on dynamical signatures of quantum chaos and information scrambling. Beyond static properties, scrambling is often diagnosed dynamically through
correlation functions that reveal how information spreads in time.
Two complementary probes have become especially influential.  
The \emph{spectral form factor} (SFF) captures correlations among energy levels
and provides a bridge between quantum dynamics and random matrix theory,
revealing universal signatures such as the dip–ramp–plateau structure associated
with chaotic spectra.  
The \emph{out-of-time-ordered correlator} (OTOC), by contrast, quantifies the
growth of operator complexity in the Heisenberg picture and measures the
sensitivity of quantum evolution to local perturbations—often interpreted as a
quantum analogue of the classical Lyapunov exponent 
\cite{Hosur2016ChaosChannels,Swingle2018UnscramblingOTOC,Schmidt2022OTOCreview}.

Finally, while exact Haar randomness is exponentially costly to realize, its
local signatures can emerge surprisingly quickly.  
Chaotic many-body dynamics and local random circuits generate approximate
Haar-like statistics on small subsystems after relatively modest evolution
times.  
This idea is formalized through unitary \(t\)-designs: small \(t\) already
captures coarse entanglement features such as purity, while higher \(t\) probes
finer spectral information.  
Unitary designs now appear throughout quantum information science—from random
circuit models and circuit complexity theory to practical protocols such as
randomized benchmarking \cite{emerson2005scalable,magesan2011rb} and
cross-entropy benchmarking \cite{boixo2018XEB,arute2019quantum}.  
In this way, the mathematics of Haar randomness and random matrix theory
provides not only a conceptual framework for understanding scrambling, but also
a set of tools that directly connect to contemporary quantum technologies, allowing benchmarking quantum processors.

The lecture notes are organized as follows:  
In Sec.~\ref{sec:closed_systems} we introduce the geometry of the unitary group \(U(D)\) as the natural configuration space for closed quantum evolution, emphasizing its Lie-algebra structure and the role of the Haar measure as the unique invariant measure on the group.  
Building on this foundation,  in Sec.~\ref{sec:Haar_measure}  we derive the induced ensemble of reduced density matrices obtained from Haar-random pure states, leading to the fixed-trace Wishart distribution, eigenvalue repulsion, the Marchenko--Pastur law, and Page’s theorem.  
We then turn to dynamical probes of quantum chaos in Sec.~\ref{sec:scrambling_dynamics}, where the spectral form factor and out-of-time-ordered correlators reveal how spectral correlations and operator growth diagnose the onset of scrambling.  Next, in Sec.~\ref{sec:t_desing} we introduce concept of untary $t$-designs approximating first $t$ moments of Haar measure.
Finally, Sec.~\ref{sec:benchmarking} connects these mathematical tools to experimentally relevant settings, explaining how unitary \(t\)-designs, random circuits, and randomized benchmarking protocols exploit Haar-like statistics to characterize the performance of modern quantum devices.  
We conclude in Sec.~\ref{sec:conclusions}.

}

{

\section{Closed Quantum System Dynamics}\label{sec:closed_systems}

A finite 
$D$-dimensional system is described by a complex Hilbert space 
$\mathcal{H} \cong \mathbb{C}^D$.  Pure states are not the 
vectors of $\mathcal{H}$ themselves, but the \emph{rays} they span: the vectors 
$|\psi\rangle$ and $e^{i\phi}|\psi\rangle$ represent the same physical 
state because all transition probabilities depend only on their relative 
phases.  After fixing the overall norm $\langle\psi|\psi\rangle = 1$, 
the space of state vectors lies on the unit sphere $S^{2D-1} \subset 
\mathbb{C}^D$, and identifying points that differ by a global phase 
$e^{i\phi} \in U(1)$ yields the manifold of physical pure states, i.e. the complex projective space,
\[
\mathcal{P}(\mathcal{H}) = S^{2D-1}/S^1 \;\cong\; \mathbb{CP}^{D-1}.
\]

All closed quantum systems evolve under unitary dynamics, and the unitary time evolution 
acts on this projective state space through one-parameter families 
$U(t) = e^{-iHt}$ generated by hermitian Hamiltonians $H$.  The unitary 
group $U(D)$ thus provides the full set of reversible transformations of 
the system: any transformation that preserves inner products and hence 
transition probabilities must be implemented by a unitary operator 
$U$ satisfying $U^\dagger U = \mathbb{I}_D$.  Consequently, the set of 
all physically admissible evolutions is precisely the unitary group 
$U(D)$.

The unitary group plays a dual role in both geometry and dynamics.
Geometrically, it is the full symmetry group of the complex Hilbert space, describing all rotations that preserve the inner product structure.
Dynamically, it provides the natural configuration space for quantum evolution:
each continuous trajectory $U(t) = e^{-iHt}$ corresponds to a path on this manifold generated by a Hermitian Hamiltonian $H$.
The tangent space of $U(D)$ thus encodes all possible infinitesimal time evolutions, while the group manifold itself represents the global space of finite quantum transformations~\cite{bengtsson2017geometry}.

Because $U(D)$ is compact, it supports a unique uniform (Haar) measure, enabling one to define averages and probability distributions over all possible unitaries.
This structure underpins the concepts of Haar-random states, random circuits, and quantum typicality that appear throughout the study of quantum chaos and information scrambling.
For this reason, a precise understanding of $U(D)$ and its Lie algebra $\mathfrak{u}(D)$ provides the natural mathematical foundation for all that follows.

\subsection{Unitary group \texorpdfstring{$U(D)$}{U(D)}}

The \emph{unitary group} \(U(D)\) is the set of all \(D\times D\) complex matrices
that preserve the Hermitian inner product on \(\mathbb{C}^D\):
\begin{equation}\label{eq:UD_definition}
U(D) = \{\, U \in \mathbb{C}^{D\times D} \mid U^\dagger U = U U^\dagger = \mathbb{I}_D \,\}.
\end{equation}
Every element \(U \in U(D)\) is invertible with \(U^{-1} = U^\dagger\), and the
group operation is matrix multiplication with identity element \(\mathbb{I}_D\).
Because every unitary matrix satisfies \(\mathrm{Tr}(U^\dagger U) = D\), all
elements share the same Frobenius norm \(\|U\|_F = \sqrt{D}\), so \(U(D)\) is a
bounded subset of \(\mathbb{C}^{D^2}\).  The defining condition
\(U^\dagger U = \mathbb{I}_D\) is closed, hence \(U(D)\) is compact.  Compactness
guarantees the existence of a unique (up to normalization) left- and
right-invariant measure on \(U(D)\): the \emph{Haar measure} (see App.~\ref{sec:haar_measure_detail}).

For compact groups the Haar measure has finite total volume and can be
normalized to unity,
\begin{equation}
\int_{U(D)} d\mu(U) = 1.
\end{equation}
This measure defines a uniform distribution over the group manifold and provides
the mathematical foundation for random unitary ensembles and Haar-random pure
states.

A closely related subgroup is the \emph{special unitary group}
\begin{equation}
SU(D) = \{\, U \in U(D) \mid \det U = 1 \,\},
\end{equation}
which excludes the overall global phase and describes “pure rotations” in
Hilbert space.  Both \(U(D)\) and \(SU(D)\) are connected Lie groups that play
central roles in quantum mechanics and gauge theory.

Physically, \(U(D)\) represents the set of all reversible quantum evolutions on
a \(D\)-dimensional Hilbert space.  The Haar measure on \(U(D)\) defines a
uniform probability distribution over such transformations and is the natural
starting point for the statistical study of quantum chaos and information
scrambling.


The unitary group \(U(D)\) is a compact Lie group of real dimension \(D^2\) \cite{Schwichtenberg2018}.
This means it is simultaneously a differentiable manifold and a group, and its local (infinitesimal) structure is captured by its Lie algebra \(\mathfrak{u}(D)\). The Lie algebra \(\mathfrak{u}(D)\) is defined as the tangent space to \(U(D)\) at the identity element \(\mathbb{I}_D\):
\begin{equation}
\mathfrak{u}(D) = T_{\mathbb{I}_D} U(D).
\end{equation}
It consists of all complex \(D\times D\) matrices \(X\) such that the exponential map
\[
U = e^{tX}
\]
remains within \(U(D)\) for all real \(t\); 
the exponential map \(U = e^X\) provides a local parametrization of the group.

Differentiating the unitarity condition \(U^\dagger U = \mathbb{I}_D\) at \(t=0\) yields
\[
\frac{d}{dt}(U^\dagger U)\big|_{t=0} = X^\dagger + X = 0,
\]
which implies
\begin{equation}
\mathfrak{u}(D) = \{\, X \in \mathbb{C}^{D\times D} \mid X^\dagger = -X \,\}.
\label{eq:lie_algebra_ud}
\end{equation}
Thus, the Lie algebra of \(U(D)\) consists of all \emph{skew-Hermitian} matrices.
The matrix exponential then provides a smooth map from the algebra to the group:
\[
\exp : \mathfrak{u}(D) \to U(D), \quad X \mapsto e^{X},
\]
which is surjective for sufficiently small neighborhoods of the identity.

It is often convenient in physics to parameterize elements of the Lie algebra using Hermitian operators.
For any Hermitian matrix \(H = H^\dagger\), the combination
\[
X = -iH
\]
belongs to \(\mathfrak{u}(D)\).
The corresponding unitary evolution
\begin{equation}
U(t) = e^{-iHt}
\label{eq:unitary_dynamics}
\end{equation}
then represents the continuous one-parameter subgroup of \(U(D)\) generated by \(H\). Equation~\eqref{eq:unitary_dynamics} is precisely the mathematical statement of quantum time evolution governed by the Schrödinger equation,
\[
i\frac{dU}{dt} = H U.
\]
Hence, Hermitian Hamiltonians correspond to elements of the tangent space of the unitary group at the identity, and the exponential map \(U = e^{-iH}\) integrates these infinitesimal generators into global unitary transformations.

At each point \(U \in U(D)\), the tangent space is obtained by translating the Lie algebra via the differential of the left (or right) group action:
\[
T_U U(D) = U \, \mathfrak{u}(D).
\]
This means any infinitesimal displacement from \(U\) within the group manifold can be written as
\[
\delta U = U X, \qquad X \in \mathfrak{u}(D).
\]
Geometrically, this corresponds to a small rotation in Hilbert space generated by a Hermitian operator \(H\), since \(X = -iH\).

}

{

\subsection{Riemannian Structure of the Unitary Group}

The unitary group \( U(D) \) is not only a Lie group but also a smooth Riemannian manifold. 
A \emph{Riemannian structure} endows a manifold with an inner product on each of its tangent spaces, 
allowing one to define angles, lengths, and distances between infinitesimal displacements---thus turning the abstract group manifold into a geometric space.

At any point \( U \in U(D) \), the tangent space is obtained by translating the Lie algebra of skew-Hermitian matrices:
\[
T_U U(D) = U\, \mathfrak{u}(D), \qquad 
\mathfrak{u}(D) = \{ X \in \mathbb{C}^{D\times D} \, | \, X^\dagger = -X \}.
\]
A natural choice of inner product on \( \mathfrak{u}(D) \) is the 
\emph{Hilbert--Schmidt inner product}
\begin{equation}
\langle X, Y \rangle = -\mathrm{Tr}(XY),
\qquad X,Y \in \mathfrak{u}(D),
\label{eq:riemannian_metric}
\end{equation}
which induces a bi-invariant Riemannian metric on the whole group. 
Bi-invariance means that left or right multiplication by any unitary does not change inner products:
\[
\langle UXU^{\dagger}, UYU^{\dagger} \rangle = \langle X, Y \rangle.
\]
This is the canonical metric on \( U(D) \) and appears throughout the theory of compact Lie groups~\cite{hall2015lie,helgason1978differential,lee2013smooth}.
In quantum mechanics, this metric gives a geometric formulation of unitary evolution and connects directly to the Fubini--Study metric on projective Hilbert space~\cite{ashtekar1999geometrical,Chu1996,brody2001geometric,nielsen2006geometric}.

Under this metric, the manifold \( U(D) \) can be viewed as a curved hypersurface embedded in \( \mathbb{C}^{D^2} \).
For a tangent vector \( X \in T_U U(D) \), its norm
\[
\|X\|^2 = -\mathrm{Tr}(X^2)
\]
measures the ``speed'' of an infinitesimal rotation in Hilbert space.
For a smooth path \( U(t) \) on the manifold, the total length is
\begin{equation}
L[U(t)] = \int_{0}^{1} \!\sqrt{ \langle \dot{U}U^{\dagger}, \dot{U}U^{\dagger} \rangle }\, dt 
= \int_{0}^{1} \!\sqrt{\mathrm{Tr}(\dot{U}^{\dagger}\dot{U})}\, dt,
\end{equation}
which is invariant under left and right translations.
The shortest such paths, called \emph{geodesics}, are generated by constant skew-Hermitian elements:
\[
U(t) = e^{-iHt}, \qquad H = H^\dagger,
\]
corresponding to uniform rotations driven by a fixed Hamiltonian \( H \).

As an example let's us consider $U(D = 2)$ group, where
every element of \( SU(2) \) can be parameterized by Euler angles,
\[
U(\alpha,\beta,\gamma,\delta) = e^{i\alpha}
\begin{pmatrix}
e^{i\beta}\cos\gamma & e^{i\delta}\sin\gamma\\
-e^{-i\delta}\sin\gamma & e^{-i\beta}\cos\gamma
\end{pmatrix}.
\]
Using the metric~\eqref{eq:riemannian_metric}, one finds the infinitesimal line element
\[
ds^2 = \mathrm{Tr}(dU^{\dagger}dU)
      = d\gamma^2 + \cos^2\gamma\, d\beta^2 + \sin^2\gamma\, d\delta^2,
\]
which corresponds to the standard metric on the three-sphere \( S^3 \), the manifold of \( SU(2) \).
Hence, the geometry of the simplest unitary group is that of a sphere of rotations, 
and Schrödinger evolution corresponds to moving along great circles on this sphere (see App.\ref{app:su2_metric}).

The correspondence between Hermitian generators and tangent vectors gives the unitary group both 
its geometric and dynamical meaning: infinitesimal displacements \( U \to U e^{-iH dt} \)
represent time evolution under the Hamiltonian \( H \).
Thus, Schrödinger dynamics can be seen as \emph{geodesic motion} on \( U(D) \) with constant ``speed'' 
\(\|H\|\) measured by the metric~\eqref{eq:riemannian_metric}~\cite{Nielsen2006}.
Such a geometric viewpoint underlies modern treatments of quantum circuit complexity, 
where distances on \( U(D) \) quantify the minimal computational ``cost'' of implementing a unitary~\cite{nielsen2006geometric}.

}

\section{The Haar Measure and Induced Ensembles}\label{sec:Haar_measure}

The Haar measure on the unitary group \(U(D)\) defines the unique probability measure \(d\mu(U)\) that is invariant under both left and right multiplication:
\begin{equation}
d\mu(U) = d\mu(VU) = d\mu(UV), \quad \forall\, V \in U(D).
\end{equation}
This invariance expresses the idea that e  pure state in the Hilbert space is equally probable under the Haar ensemble.  
Sampling a unitary \(U\) from the Haar measure corresponds to drawing a uniformly random rotation in Hilbert space (see App.\ref{sec:haar_measure_detail}).

Let a composite system be described by the Hilbert space
$\mathcal{H} = \mathcal{H}_A \otimes \mathcal{H}_B$,
where subsystem \(A\) has dimension \(m = \dim \mathcal{H}_A\) and subsystem \(B\) has dimension \(n = \dim \mathcal{H}_B\), with total dimension \(D = mn\).
We begin from a reference pure state \(\ket{\psi_0}\) and apply a Haar-random unitary \(U \in U(D)\):
\begin{equation}
\ket{\psi} = U \ket{\psi_0}.
\end{equation}
The state \(\ket{\psi}\) is therefore uniformly distributed over the unit sphere in \(\mathbb{C}^D\).

{ 
\subsection{Reduced Density Matrices and Induced Measures}\label{sec:rdm_haar}

Dividing the system into subsystems \(A\) and \(B\) with Hilbert spaces
\(\mathcal{H}_A\) and \(\mathcal{H}_B\) of dimensions \(m\) and \(n\) respectively,
we define the reduced density matrix of \(A\) by tracing over the environment:
\begin{equation}
\rho_A = \mathrm{Tr}_B \ket{\psi}\bra{\psi}.
\end{equation}
For a pure state \(\ket{\psi}\) chosen uniformly with respect to the Haar measure on the total Hilbert space \(\mathcal{H}_A \otimes \mathcal{H}_B\),
the reduced state \(\rho_A\) becomes a random mixed state.
The statistics of such random mixed states are \emph{induced} by the Haar measure on the full system~\cite{zyczkowski2001induced,zyczkowski2003hilbert,bengtsson2017geometry}.

A Haar-random pure state can be written in a product basis as
\begin{equation}
\ket{\psi} = \sum_{i=1}^{m}\sum_{\mu=1}^{n} C_{i\mu}\,
\ket{i}_A \otimes \ket{\mu}_B,
\end{equation}
where the complex coefficients \(C_{i\mu}\) satisfy the normalization
\(\sum_{i,\mu} |C_{i\mu}|^2 = 1.\)
The joint distribution of the coefficients follows from the unitary invariance of the Haar measure:
\begin{equation}
P(\{C_{i\mu}\}) \propto
\delta\!\left(1 - \sum_{i,\mu} |C_{i\mu}|^2\right).
\label{eq:haar_joint}
\end{equation}
This uniform distribution on the complex unit sphere in \(\mathbb{C}^{mn}\) is equivalent to drawing an unnormalized complex Gaussian random vector and normalizing it.
Hence, the coefficients behave as independent complex Gaussian variables with zero mean and variance \(1/(mn)\)~\cite{hayden2006aspects,nadal2010statistical}.
Equivalently, the matrix \(C\) can be viewed as a random matrix from the complex Ginibre ensemble with an overall normalization constraint (see App.\ref{app:haar_gaussian}).

The reduced density matrix of subsystem \(A\) reads
\begin{align}
\rho_A &= \mathrm{Tr}_B \ket{\psi}\bra{\psi} \nonumber \\
&= \sum_{i,j=1}^{m}\left(\sum_{\mu=1}^{n} C_{i\mu} C_{j\mu}^* \right)\ket{i}\bra{j}.
\end{align}
Defining an \(m \times n\) matrix \(X\) with entries \(X_{i\mu} = C_{i\mu}\sqrt{mn}\),
we can write this compactly as
\begin{equation}
\rho_A = \frac{X X^{\dagger}}{\mathrm{Tr}(X X^{\dagger})}.
\label{eq:rhoA}
\end{equation}
This expression explicitly shows that \(\rho_A\) is proportional to a random positive semidefinite matrix formed by multiplying a Gaussian matrix with its Hermitian conjugate.

The unnormalized matrix $W = X X^{\dagger}$
is a positive semidefinite Hermitian matrix belonging to the \emph{complex Wishart ensemble}.
If the entries of \(X\) are independent complex Gaussian variables of unit variance, then \(W\) is distributed as~\cite{wishart1928generalised,Forrester1993,mehta2004random,zyczkowski2001induced,nadal2010statistical}:
\begin{equation}
P(W)\,dW \propto (\det W)^{n - m} e^{-\mathrm{Tr}\, W}\,dW,
\label{eq:wishart}
\end{equation}
where \(dW\) denotes the flat measure on Hermitian matrices.
The Wishart ensemble was first introduced in statistics as the distribution of sample covariance matrices \cite{wishart1928generalised,Aitken1949}, but in quantum information it appears naturally when tracing over subsystems of Haar-random states.

The normalized reduced density matrix \(\rho_A\) defined in Eq.~\eqref{eq:rhoA}
is obtained from the Wishart matrix \(W\) by imposing the normalization constraint \(\mathrm{Tr}\,W = 1\).
The resulting ensemble of density matrices is therefore a \emph{fixed-trace Wishart ensemble}~\cite{zyczkowski2001induced,nadal2010statistical,bengtsson2017geometry}.
Its probability measure on the space of density operators takes the form:
\begin{equation}
P(\rho_A)\,d\rho_A
\propto
(\det \rho_A)^{n - m}
\, \delta(\mathrm{Tr}\,\rho_A - 1)\,d\rho_A.
\label{eq:induced}
\end{equation}

The induced measure constitutes the natural statistical framework for analyzing the properties of subsystems of Haar-random pure states.
In this formulation, tracing out environmental degrees of freedom transforms a Haar-random state into a density operator drawn from the fixed-trace Wishart ensemble.
}

\subsection{Eigenvalue Statistics and Joint Distribution}

Having established that $\rho_A$ follows the fixed-trace Wishart distribution, we now turn to its eigenvalues $\{\lambda_i\}_{i=1}^{m}$, which encode the full entanglement spectrum of subsystem $A$.
By definition, these eigenvalues are non-negative and satisfy $\sum_i \lambda_i = 1$.
They quantify how mixed the subsystem appears when the global state is pure.

The change of variables from the matrix entries of $W = X X^{\dagger}$ to its eigenvalues and eigenvectors introduces a Jacobian that reflects the curvature of the space of Hermitian matrices.
Integrating out the unitary degrees of freedom of $W$, one obtains the celebrated joint probability density for the eigenvalues~\cite{zyczkowski2001induced,nadal2010statistical}:
\begin{equation}
P(\lambda_1, \ldots, \lambda_m)
\propto
\delta\!\left(\sum_{i=1}^m \lambda_i - 1\right)
\prod_{i=1}^{m}\lambda_i^{n - m}
\prod_{i < j} (\lambda_i - \lambda_j)^2.
\label{eq:joint}
\end{equation}

The factor $\prod_{i < j} (\lambda_i - \lambda_j)^2$ is the \emph{Vandermonde determinant squared}, characteristic of unitary-invariant random matrix ensembles.  
It encodes \emph{level repulsion}: eigenvalues cannot coincide because the measure of such configurations vanishes.
This is the statistical signature of the repulsion between the eigenvalues of $\rho_A$, analogous to the repulsion of energy levels in quantum chaotic systems (see  App.\ref{app:RMT_intro}, App.\ref{app:eig_dist}).

In the context of quantum information, eigenvalue repulsion translates into the fact that entanglement is not dominated by a few large Schmidt coefficients but is evenly spread across the subsystem’s Hilbert space.  
This statistical uniformity implies that random pure states exhibit nearly maximal entanglement entropy and are locally indistinguishable from thermal states.
Such eigenvalue rigidity is a central feature of chaotic dynamics and is absent in integrable or many-body localized systems, where the entanglement spectrum becomes sparse and non-repelling.

The joint eigenvalue distribution, Eq.~\eqref{eq:joint}, shows that tracing over a large environment produces a spectrum with universal level repulsion, reflecting the uniform spreading of quantum information across the Hilbert space.
Any quantity that depends on the spectrum of $\rho_A$ can be computed by averaging over~\eqref{eq:joint}.
Typical examples include the $k$th spectral moment $\mathbb{E}[\mathrm{Tr}(\rho_A^k)]$ and the entanglement entropy $S(\rho_A) = -\mathrm{Tr}(\rho_A \log \rho_A)$.
The simplicity of Eq.~\eqref{eq:joint} allows for analytical computation of the first few moments, leading to expressions for purity and entropy that reveal how small subsystems of Haar-random states appear nearly thermal.

\subsection{Spectral Moments and Purity}

The eigenvalues $\{\lambda_i\}$ of $\rho_A$ encode all local statistical information accessible to subsystem $A$.  
Moments of the spectrum,
\begin{equation}
M_k = \mathbb{E}\!\left[\mathrm{Tr}(\rho_A^k)\right] = \mathbb{E}\!\left[\sum_{i=1}^{m}\lambda_i^k\right],
\end{equation}
quantify how mixed the subsystem is.  
The second moment $M_2$ corresponds to the \emph{purity} $\gamma = \mathrm{Tr}(\rho_A^2)$, while $S(\rho_A) = -\mathrm{Tr}(\rho_A \log \rho_A)$ measures the entropy.  
Higher moments describe finer features of the entanglement spectrum.

Normalization of the density matrix implies
\begin{equation}
\sum_{i=1}^{m}\lambda_i = 1 \quad \Rightarrow \quad \mathbb{E}[\lambda] = \frac{1}{m}.
\label{eq:first_moment}
\end{equation}
This simple result is exact for all $m,n$ and reflects that, on average, each eigenvalue contributes equally to the unit trace of $\rho_A$.  
It already signals that a typical subsystem is close to maximally mixed.

The second spectral moment can be evaluated by integrating Eq.~\eqref{eq:joint}, yielding~\cite{zyczkowski2001induced}:
\begin{equation}
\mathbb{E}[\lambda^2] = \frac{m + n}{m(mn + 1)}.
\label{eq:second_moment}
\end{equation}
From this, the variance of a single eigenvalue follows:
\begin{equation}
\mathrm{Var}(\lambda) = \frac{m^2 - 1}{m^2(mn + 1)}.
\label{eq:variance}
\end{equation}
In the limit $n \gg m$, $\mathrm{Var}(\lambda) \sim 1/n$, showing that as the environment grows, eigenvalues become increasingly uniform—an explicit manifestation of typicality.

The purity of the subsystem is defined by
\begin{equation}
\gamma = \mathrm{Tr}(\rho_A^2) = \sum_{i=1}^m \lambda_i^2.
\end{equation}
Averaging over the Haar ensemble gives~\cite{page1993average}:
\begin{equation}
\mathbb{E}[\gamma] = \frac{m + n}{mn + 1}.
\label{eq:purity}
\end{equation}
For large $n$,
\begin{equation}
\mathbb{E}[\gamma] \simeq \frac{1}{m} + \frac{1}{n} + \mathcal{O}\!\left(\frac{1}{n^2}\right),
\end{equation}
which approaches $1/m$, the purity of a maximally mixed state on $m$ levels.  
This scaling illustrates how subsystems of large entangled systems appear thermal:  
the effective state on $A$ is nearly maximally mixed, with deviations suppressed by $1/n$.

Equation~\eqref{eq:purity} quantifies the central idea of \emph{typicality}:  
although the global state is pure, the reduced density matrix of a small subsystem behaves as if it were thermal.  
This is a direct statistical consequence of Hilbert-space geometry rather than any specific dynamical mechanism (see App.\ref{sec:typicality}).

In the language of many-body physics, this result provides the static foundation for the \emph{Eigenstate Thermalization Hypothesis} (ETH)~\cite{deutsch1991quantum,srednicki1994chaos}, which asserts that individual energy eigenstates of chaotic Hamiltonians appear thermal when observed locally.
Purity decays inversely with environment size, showing quantitatively how entanglement between subsystem and environment enforces apparent thermalization.

\subsection{Asymptotic Spectrum: The Marchenko--Pastur Law}

When both subsystems $A$ and $B$ are large, the discrete eigenvalues $\{\lambda_i\}$ of $\rho_A$ form a dense spectrum that can be described by a continuous probability density $P(\lambda)$.  
In the limit $m,n \to \infty$ with a fixed ratio $c = m/n \le 1$, the distribution of eigenvalues converges to the celebrated \emph{Marchenko--Pastur (MP) law}~\cite{marchenko1967distribution}:
\begin{equation}
P(\lambda)
= \frac{mn}{2\pi}
  \frac{\sqrt{(\lambda_+ - \lambda)(\lambda - \lambda_-)}}{\lambda},
  \qquad \lambda \in [\lambda_-, \lambda_+],
\label{eq:MP}
\end{equation}
where the spectral edges are given by
\begin{equation}
\lambda_\pm = \frac{1}{m}\left(1 \pm \sqrt{\frac{m}{n}}\right)^2.
\label{eq:edges}
\end{equation}
Outside this interval, $P(\lambda) = 0$, and the normalization condition $\int_{\lambda_-}^{\lambda_+} P(\lambda)\,d\lambda = 1$ holds exactly.

The Marchenko--Pastur distribution characterizes the eigenvalue spectra of random covariance matrices and, by extension, reduced density matrices of Haar-random pure states. It shows that most eigenvalues lie close to $1/m$, indicating that $\rho_A$ is nearly maximally mixed, the width of the distribution, $\lambda_+ - \lambda_-$, scales as $\sqrt{m/n}/m$, shrinking as the environment grows. 
The MP law depends only on $c = m/n$ and not on microscopic details—an archetypal example of universality in random matrix theory.

Near the spectral edges $\lambda_{\pm}$, the density vanishes as $\sqrt{|\lambda - \lambda_{\pm}|}$, exhibiting a universal square-root singularity.
For finite but large $m,n$, the largest and smallest eigenvalues fluctuate around $\lambda_{\pm}$ on a scale $\sim m^{-2/3}$ and follow the \emph{Tracy--Widom distribution}~\cite{tracy1994level,tracy1996orthogonal,Forrester1993}.

Eq.~\eqref{eq:MP} implies that the entanglement spectrum of a small subsystem in a random pure state mimics the eigenvalue spectrum of a thermal density matrix.  
The degree of deviation from maximal mixing decreases rapidly as the subsystem becomes smaller relative to the environment.  
This is the quantitative statement of \emph{local thermalization} in the absence of a heat bath: typical entanglement acts as an effective thermodynamic reservoir.

\subsection{Entanglement Entropy and Page's Theorem}

The von Neumann entropy of a subsystem quantifies its entanglement with the environment.  
For a pure bipartite state $\ket{\psi}_{AB}$, the reduced density matrix $\rho_A = \Tr_B \ket{\psi}\bra{\psi}$ has entropy
\begin{equation}
S(\rho_A) = -\Tr(\rho_A \log \rho_A).
\label{eq:entropy}
\end{equation}
If $\rho_A$ were maximally mixed, all eigenvalues would be equal to $1/m$, and the entropy would reach its maximal value $S_{\max} = \log m$.  
Deviations from maximality reflect residual purity due to the finite environment dimension $n$.

In the context of scrambling, $S(\rho_A)$ measures how completely information initially localized in subsystem $A$ becomes spread across the system.
A Haar-random state represents the endpoint of perfect scrambling, and its entanglement entropy provides a statistical benchmark for chaotic systems.

In Ref.\cite{page1993average}, Page derived the exact average entropy of a subsystem of dimension $m$ coupled to an environment of dimension $n \ge m$ when the total state is Haar-random~\cite{page1993average}:
\begin{equation}
\mathbb{E}[S(\rho_A)] = \sum_{k = n+1}^{mn} \frac{1}{k} - \frac{m - 1}{2n}.
\label{eq:page_exact}
\end{equation}
For large $m,n$, this simplifies to
\begin{equation}
\mathbb{E}[S(\rho_A)]
\simeq
\log m - \frac{m}{2n} + \mathcal{O}\!\left(\frac{1}{n^2}\right).
\label{eq:page_asym}
\end{equation}
Equation~\eqref{eq:page_asym} shows that $\mathbb{E}[S]$ approaches $\log m$, with a small correction reflecting the finite-size deviation from maximal mixing. Page's theorem quantifies how Haar-randomness leads to nearly maximal entanglement.
It provides the statistical ceiling of subsystem entropy, serving as a benchmark for information scrambling in both quantum circuits and black holes.

When the total system has dimension $D = mn$ and is partitioned into two complementary subsystems, the average entanglement entropy of one subsystem depends only on its Hilbert-space dimension.
The function $S_A(m)$ is symmetric under $m \leftrightarrow n$, rising with $m$ until $m \approx n$ and then saturating.
This behavior is known as the \emph{Page curve}, and it captures the transition from low to maximal entanglement.

Page's theorem formalizes the idea that almost e  pure state in a large Hilbert space is locally thermal.  
The reduced state $\rho_A$ is nearly indistinguishable from the maximally mixed state $\mathbb{I}_A/m$, and deviations scale as $\mathcal{O}(1/n)$.
This is a purely kinematic property of high-dimensional Hilbert spaces, independent of dynamics.

In dynamical systems, such as many-body Hamiltonians or random circuits, the approach of $S(\rho_A)$ toward the Page value defines the \emph{scrambling time}---the time required for the system to reach local typicality.
In strongly chaotic systems, this time scales logarithmically with the total system size~\cite{Sekino2008,lashkari2013towards}.

\section{Dynamical diagnostics of chaos and scrambling}\label{sec:scrambling_dynamics}

So far, we have characterized the \emph{static} spectral properties of reduced density matrices drawn from the Haar ensemble.  
The static properties of Haar-random states and induced ensembles capture the
entanglement structure of fully scrambled quantum states.  To understand how
scrambling emerges dynamically, we now study how quantum correlations evolve
under unitary time evolution.
Two
complementary diagnostics play central roles:
\begin{itemize}
  \item The \emph{spectral form factor} (SFF) probes correlations among
        energy eigenvalues and reveals the onset of universal random-matrix
        behavior in the spectrum.

  \item The \emph{out-of-time-ordered correlator} (OTOC) probes the
        spreading of operators in the Heisenberg picture and quantifies the
        dynamical loss of locality underlying information scrambling. OTOCs  provide
a microscopic, operator-based description of srcambling.
\end{itemize}

{
 
\subsection{Spectral form factor}
\label{sec:SFF}

A central signature of quantum chaos lies in the correlations of Hamiltonian the energy
spectrum.  Chaotic many-body systems exhibit level repulsion and long-range
spectral rigidity, in contrast to integrable models, whose spectra are
characterized by uncorrelated (Poissonian) energy levels (for introduction to Random Matrix Theory see Appendix~\ref{app:RMT_intro}).  The \emph{spectral
form factor} (SFF) provides a compact way to isolate these correlations and
connect them to real-time dynamics.

Let us consider system defined by Hamiltonian $H$ living in $D$ dimensional Hilbert space, $H\ket{n}=E_n\ket{n}$.  Introducing the complex-time partition function
\begin{equation}
  Z(\beta+it)
  =
  \Tr\!\left[e^{-(\beta+it)H}\right]
  = \sum_{n=1}^{D} e^{-(\beta+it)E_n},
\end{equation}
the spectral form factor is defined as
\begin{equation}\label{eq:SFF-def}
  K(\beta,t)=\big|Z(\beta+it)\big|^{2} =
 \sum_{m,n}
      e^{-\beta(E_m+E_n)} e^{-it(E_m-E_n)}.
\end{equation}
Setting $\beta=0$ yields a purely spectral diagnostic.
If the eigenvalues \( \{E_n\} \) were uncorrelated, the oscillatory phases
\( e^{-it(E_m-E_n)} \) would cancel and \( K(t) \) would decay rapidly.
Chaotic spectra behave differently.  Level repulsion suppresses small spacings,
and spectral rigidity suppresses long-wavelength fluctuations. As such, the SFF measures the self-correlation of the time-evolution operator:
how strongly the dynamics at time \( t \) interferes with itself.
Chaotic spectra suppress this overlap at intermediate times, producing the dip
and the ramp, while long-time coherence arising from spectral rigidity leads to
the plateau. This dip–ramp–plateau pattern is a robust signature of quantum chaotic dynamics~\cite{haake2010quantum}.

Integrable models  possess
extensive sets of conserved quantities and exhibit Poissonian level statistics;
accordingly, their SFF lacks a ramp
\cite{DAlessioRigol2016,Abanin2019MBLreview}.  
Adding integrability-breaking perturbations---e.g.\ disorder or
next-nearest-neighbour interactions—leads to level repulsion and the emergence
of a ramp in $K(t)$ \cite{Bertini2018}.

}

\subsection{Out-of-Time-Ordered Correlators and the Spread of Information}
\label{sec:OTOC}

Out-of-time-ordered correlators (OTOCs) provide a microscopic, dynamical
diagnostic of information scrambling in quantum many-body systems. They measure
how a local perturbation grows under unitary time evolution 
\cite{ShenkerStanford2014,hosur2016chaos,
Swingle2018UnscramblingOTOC,Swingle2018Boulder,XuSwingle2024ScramblingOTOC}.

Let $V$ and $W$ be unitary operators supported initially on disjoint spatial
regions of a many-body system with Hamiltonian $H$.  The standard OTOC is
\begin{equation}
  F(t)
  =
  \big\langle
     W^\dagger(t)\, V^\dagger \, W(t)\, V
  \big\rangle,
  \qquad
  W(t)=U^\dagger(t)W U(t),\ \ U(t)=e^{-iHt},
  \label{eq:OTOC-def}
\end{equation}
where the expectation value is taken in a reference state or thermal ensemble.
At $t=0$ one has $[W(0),V]=0$ and therefore $F(0)=1$.  As time evolves,
$W(t)$ becomes a more complicated operator with support spreading away from its
initial location.  Once $W(t)$ develops nontrivial overlap with the support of
$V$, the commutator becomes nonzero and $F(t)$ departs from~1.  This departure
is the signature of scrambling: the loss of locality in the Heisenberg picture.

A convenient companion observable is the squared commutator
\begin{equation}
  C(t)
  =
  \big\langle
     [W(t),V]^\dagger [W(t),V]
  \big\rangle
  \;\ge 0.
  \label{eq:C-def}
\end{equation}
For unitary $V$ and $W$, an elementary identity gives
\begin{equation}
  C(t) = 2\bigl(1-\mathrm{Re}\,F(t)\bigr).
  \label{eq:C-F-identity}
\end{equation}

In a lattice system with local
Hamiltonian $H=\sum_x h_x$ with finite-range interactions, Lieb--Robinson
theory \cite{lieb1972finite,bravyi2006lieb} shows that commutators between
operators with spatial separation $|x-y|$ obey
\begin{equation}
  \big\| [O_x(t), O_y] \big\|
  \;\le\;
  c\, e^{-(|x-y|-v_{\mathrm{LR}} t)/\xi},
  \label{eq:LR-bound}
\end{equation}
where $v_{\mathrm{LR}}$ is the Lieb--Robinson velocity, $\xi$ a correlation
length, and $c$ a constant depending on microscopic details.  This bound implies
a linear light-cone structure: $W(t)$ cannot develop significant support beyond
distance $v_{\mathrm{LR}}t$.

In chaotic systems, OTOCs sharpen this picture. Consider $V$ supported at
position $x$ and $W$ at $y$.  One typically observes
\begin{equation}
  C(x,y;t)
  \;\equiv\;
  \big\langle [W(y,t),V(x)]^\dagger [W(y,t),V(x)] \big\rangle
  \;\sim\;
  \exp\!\bigl[ 2\lambda_L\, (t - |x-y|/v_B) \bigr],
\end{equation}
where $\lambda_L$ is a many-body Lapunov exponent, and $v_B$ is so-called 
\emph{butterfly velocity}, determining the speed at which the operator front
moves spatially.  In local systems one always finds
\begin{equation}
  v_B \le v_{\mathrm{LR}},
\end{equation}
because the Lieb--Robinson bound limits the maximal speed at which the support
of any Heisenberg operator can expand. For lattice systems with local interactions, operator growth is
\emph{ballistic}: the spatial radius of $W(t)$ grows as $r(t)\approx v_B t$.

Inside the butterfly cone, many chaotic systems exhibit an intermediate window
in which the squared commutator grows exponentially,
\begin{equation}
  C(t)\sim \varepsilon\, e^{2\lambda_L t},
  \label{eq:C-exp}
\end{equation}
where $\lambda_L$ is a many-body Lyapunov exponent.  This exponential regime
holds until $C(t)$ becomes $\mathcal{O}(1)$, after which the OTOC saturates and
no longer carries information about exponential growth.

To estimate the \emph{scrambling time} we associate the support of $W(t)$ with
the number of microscopic degrees of freedom it acts on.  During the exponential
growth window this size behaves as $\sim e^{\lambda_L t}$.  If the system has
$N$ effective local degrees of freedom (e.g.\ $N$ spins), scrambling occurs when
the perturbation influences $\mathcal{O}(N)$ sites, i.e.
\begin{equation}
  e^{\lambda_L t_*} \sim N
  \qquad\Longrightarrow\qquad
  t_* \sim \frac{1}{\lambda_L}\,\log N.
  \label{eq:tstar-logN}
\end{equation}
In a local $d$-dimensional lattice, however, growth is limited to the ball
of radius $v_B t$, so the affected volume scales as $(v_B t)^d$.  Setting this
equal to $N$ gives $t_* \sim N^{1/d}$,
parametrically larger than~\eqref{eq:tstar-logN}.  Fast scrambling therefore
requires either all-to-all interactions or sufficiently long-range connectivity.
For systems without an obvious notion of ``sites'', it is natural to measure the
system size by its Hilbert-space dimension $D$.  The entropy
$S=\log D$ then quantifies the number of orthogonal states accessible to the
system.  Replacing $N$ by $S$ in~\eqref{eq:tstar-logN} gives
\begin{equation}
  t_* \sim \frac{1}{\lambda_L}\, \log S,
  \label{eq:tstar-S}
\end{equation}
a useful form in quantum field theory and gravitational contexts.

The Lapunov exponent $\lambda_L$ is an upper bounded quantity~\cite{Maldacena2016}, i.e. for a thermal quantum systems at temperature $T=1/\beta$ the following takes place
\begin{equation}
  \lambda_L \le 2\pi T.
  \label{eq:chaos-bound}
\end{equation}
Systems saturating this bond,~\eqref{eq:chaos-bound}, are called \emph{maximally chaotic}. Example of such a system is given by black holes. Expressing scrambling time via notion of entropy, allows connect scrambling dynamics with black holes.
A conceptual realization of maximally chaotic system is given by black holes~\cite{lloyd1988black, hayden2007black}. Their number of microscopic degrees
of freedom is encoded in the Bekenstein--Hawking entropy
\begin{equation}
  S_{\mathrm{BH}} = \frac{A}{4G\hbar},
\end{equation}
with $A$ the horizon area~\cite{Bekenstein1973BHentropy,Hawking1975}.  Assuming
saturation of the Lapunov exponent, and using~\eqref{eq:tstar-S} yields
\begin{equation}
  t_*^{\mathrm{BH}}
  \sim \frac{1}{2\pi T}\,\log S_{\mathrm{BH}},
\end{equation}
which grows only logarithmically with entropy.  Black holes therefore behave as
\emph{fast scramblers}~\cite{HaydenPreskill2007,Sekino2008}.

A microscopic example of maximally chaotic quantum states is given by the Sachdev--Ye--Kitaev (SYK)
model~\cite{SachdevYe1993GaplessSpinFluid,MaldacenaStanford2016CommentsSYK}.
In its complex-fermion form it consists of $N$ fermions with all-to-all random
four-fermion couplings,
\begin{equation}
  H_{\mathrm{SYK}}
  = \sum_{i,j,k,\ell=1}^N J_{ij;k\ell}\,
    c_i^\dagger c_j^\dagger c_k c_\ell,
\end{equation}
where $J_{ij;k\ell}$ are antisymmetric Gaussian couplings with variance
$\overline{|J_{ij;k\ell}|^2}=J^2/N^3$.  Because every fermion interacts with 
every other, $W(t)$ spreads over all $N$ sites at an exponential rate, giving
$t_* \sim (1/\lambda_L)\log N$ and saturating the chaos bound at finite
temperature.  Experimental analogues of SYK dynamics have been proposed in modern quantum simulators platforms~\cite{Danshita2017SYKcoldatoms,Wei2021,Landsman2019IonOTOC,Chew2017GrapheneSYK}.

Random circuits offer a complementary perspective.  In local architectures each
gate acts on nearby qubits, so operator growth is restricted by a light cone
with velocity $v_B$ set by the circuit structure.  The operator front expands
ballistically and the affected volume grows as $(v_B t)^d$, implying
$t_* \sim N^{1/d}$ and therefore slow scrambling
\cite{nahum2018operator,chan2018spectral,zhang2019scrambling}.  By contrast,
circuits with long-range gates or all-to-all connectivity display exponential
operator growth and therefore logarithmic scrambling
\cite{Bertini2020,lashkari2013towards,haferkamp2022linear}.  Such
circuits are algorithmic realizations of the fast-scrambling phenomenology
encoded in OTOCs.

\section{Haar-Random States and Their Physical Approximations}\label{sec:t_desing}

Information scrambling is deeply intertwined with computational complexity:
chaotic dynamics rapidly generate highly entangled states whose structure becomes
exponentially hard to distinguish from Haar-random states%
~\cite{susskind2016computational,brown2018secondlaw,roberts2017chaoscomplexity,haferkamp2022linear,Chapman2022}.
This connection places scrambling at the interface of quantum chaos and the theory
of quantum circuits.  
Haar randomness represents an idealized limit of maximal entanglement, maximal
complexity, and complete loss of local information.

In realistic settings, however, exact Haar randomness is exponentially costly to
achieve.  Physical systems instead approximate random behavior through chaotic
time evolution or through local quantum circuits of finite depth.  Such dynamics
can reproduce Haar-like statistics on small subsystems after modest evolution
times.  A convenient language for formalizing this approximation is provided by
\emph{unitary $t$-designs}~\cite{dankert2009exact,brandao2016local,Harrow2023-xd,hunterjones2019unitary},
ensembles of unitaries that agree with Haar measure up to the first \(t\) moments.

To state this precisely, it is useful to introduce the notion of \emph{twirling}.
Given a probability measure \(\mu\) on the unitary group \(\mathrm{U}(D)\), the
associated twirling channel is the conjugation average
\begin{equation}
    \mathcal{T}^{(\mu)}(X)
    = \int_{\mathrm{U}(D)} \mathrm{d}\mu(U)\,
        U X U^\dagger ,
\end{equation}
which maps any operator \(X\) to its symmetrized version under the action of
\(\mu\).  When \(\mu\) is the Haar measure, \(\mathcal{T}_{\mu}\) projects onto
operators that are invariant under all unitaries; this mechanism underlies many
universality results in random matrix theory and quantum information.

A probability distribution $\nu$ on $\mathrm{U}(D)$ is called a (unitary) $t$-design
if its $t$-fold twirl reproduces the Haar twirl exactly:
\begin{equation}
  \int_{\mathrm{U}(D)} d\nu(U)\,
  U^{\otimes t} X U^{\dagger \otimes t}
  =
  \int_{\mathrm{U}(D)} d\mu(U)\,
  U^{\otimes t} X U^{\dagger \otimes t}
  \qquad \forall\, X ,
  \label{eq:t-design}
\end{equation}
where $\mu$ denotes the Haar measure on $\mathrm{U}(D)$.
We refer to the map
\begin{equation}
  \mathcal{T}^{(\nu)}_t(X)
  :=
  \int_{\mathrm{U}(D)} d\nu(U)\,
  \bigl( U^{\otimes t} X U^{\dagger \otimes t} \bigr)
  \qquad \forall\, X
  \label{eq:t-fold-twirl}
\end{equation}
as the $t$-fold twirl with respect to $\nu$.
As such, all degree-\(t\) polynomial moments in the matrix elements of
\(U\) agree with those of a Haar-random unitary.  Approximate \(t\)-designs
relax this equality to hold up to a small error in an appropriate operator norm.

For \(t=2\), this condition already guarantees that all quantities depending only
on second moments---such as subsystem purities, Rényi-2 entropies, and the Page
curve---are indistinguishable from those obtained from exact Haar randomness.
Higher \(t\) capture increasingly refined features of chaotic dynamics,
including fine-grained spectral correlations and temporal diagnostics encoded in
the spectral form factor \(K(t)\).  Thus, unitary designs provide a powerful
bridge between exact mathematical randomness and the practical forms of
pseudo-randomness generated by physical chaotic systems.

An  approximate
unitary 2-designs can be generated efficiently using local random quantum
circuits with a brick-wall architecture.  In such circuits, two-qubit gates are
applied in alternating even and odd layers so that every qubit interacts with
its neighbors over \(O(1)\) consecutive layers.  It is by now well established
that random circuits of depth \(O(n)\) on \(n\) qubits form an \(\varepsilon\)-approximate
2-design~\cite{dankert2009exact,brandao2016local,hunterjones2019unitary}.
Consequently, the entanglement and purity statistics of Haar-random states—
including Page’s theorem—can be reproduced efficiently by physically reasonable
quantum circuits, providing a concrete dynamical mechanism for the emergence of
Haar-like randomness in chaotic quantum systems.

{
\section{Haar Randomness and Benchmarking of Quantum Devices}
\label{sec:benchmarking}

The statistical structure of Haar-random states and unitary designs also plays a
central role in the \emph{characterization of quantum processors}.  
Because real devices are noisy and implement imperfect gates, one needs
protocols that can quantify how closely their operations approximate ideal
unitary evolution.  
Randomized methods based on approximate unitary designs provide an elegant and
experimentally practical way to do so.

A key idea connecting these protocols to Haar-random theory is the notion of a
\emph{twirl}.  
Given an ensemble of unitaries $\mathcal{E} \subset U(D)$ and a noisy quantum
operation $\mathcal{E}_{\mathrm{noise}}$, one defines the twirled channel as
\begin{equation}
  \Lambda_{\mathcal{E}}(\rho)
  = \mathbb{E}_{U\in\mathcal{E}}
  \bigl[ U^{\dagger}\,
  \mathcal{E}_{\mathrm{noise}}(U \rho U^{\dagger})\,U
  \bigr].
  \label{eq:twirl_benchmark}
\end{equation}
If the ensemble $\mathcal{E}$ forms a unitary 2-design---as is the case for the
Clifford group---this average symmetrizes any noise process into an isotropic
\emph{depolarizing channel}
\[
\Lambda_{\mathcal{E}}(\rho)
= p\,\rho + (1-p)\frac{I}{D},
\]
where \(p\) quantifies the average coherence retained per gate.
This observation forms the theoretical foundation of \emph{randomized
benchmarking} (RB)~\cite{emerson2005scalable,emerson2007symmetrized,magesan2011rb,magesan2012rb,Wallman2014}.

In RB, a device is subjected to sequences of randomly chosen Clifford gates,
each drawn from a 2-design, followed by a final gate that inverts the sequence
so that an ideal implementation returns the system to its initial state.
Averaging over many random sequences of length \(m\) yields a characteristic
exponential decay of the measured survival probability,
\begin{equation}
  F(m) = A p^{m} + B ,
\end{equation}
where \(A\) and \(B\) absorb state-preparation and measurement (SPAM) errors.
The decay constant \(p\) is directly related to the \emph{average gate fidelity}
via
\[
\mathcal{F}_{\mathrm{avg}} = \frac{(D-1)p + 1}{D}.
\]
Because 2-designs reproduce the second moments of the Haar measure, RB
effectively performs a Haar twirl up to second order, ensuring that only the
isotropic component of the noise contributes to the measured decay.  
This makes RB highly robust and scalable, allowing accurate fidelity estimates
without the need for full process tomography.

A complementary and conceptually distinct method is
\emph{cross-entropy benchmarking} (XEB),
which probes how closely a device reproduces the output statistics of a
pseudo-random circuit~\cite{boixo2018XEB,arute2019quantum}.
Rather than studying how noise accumulates during repeated gate applications,
XEB evaluates how Haar-like the distribution of measured bitstrings is.
In this protocol, one samples bitstrings \(x\) from the experimental output
distribution \(P_{\mathrm{exp}}(x)\) and compares them with the ideal
probabilities \(P_{\mathrm{ideal}}(x)\) computed for the same circuit on a
classical simulator (for as large a system as feasible).
The agreement is quantified by the \emph{cross-entropy fidelity}
\begin{equation}
  F_{\mathrm{XEB}}
  = 2^n \sum_{x} P_{\mathrm{ideal}}(x) P_{\mathrm{exp}}(x) - 1 ,
  \label{eq:fxeb}
\end{equation}
where \(n\) is the number of qubits.
This quantity takes the value \(F_{\mathrm{XEB}}=1\) for a perfectly ideal circuit
and \(F_{\mathrm{XEB}}=0\) for a fully random (uniform) or completely
depolarized output.

The interpretation of Eq.~\eqref{eq:fxeb} is rooted in the statistics of
Haar-random quantum states.  
For a truly Haar-distributed unitary acting on \(n\) qubits, the output
probabilities \(P_x = |\psi_x|^2\) follow the \emph{Porter--Thomas distribution}
first introduced in nuclear spectroscopy~\cite{porter1956fluctuations} and
later generalized in random matrix theory~\cite{mehta2004random}:
\begin{equation}
  P(P_x) = 2^n e^{-2^n P_x},
  \label{eq:porter_thomas}
\end{equation}
an exponential distribution with mean \(2^{-n}\) and variance \(2^{-2n}\).
This law encapsulates the intuition that the amplitudes of a Haar-random state
are statistically independent complex Gaussians constrained by normalization.
When a physical circuit generates states that are close to Haar-random,
the experimentally observed probabilities \(P_{\mathrm{exp}}(x)\)
exhibit this same exponential form, and the cross-entropy fidelity
\(F_{\mathrm{XEB}}\) approaches unity.  
Departures from the Porter--Thomas statistics therefore provide a quantitative
measure of residual coherence, correlated noise, or insufficient circuit depth.

Both RB and XEB exploit the statistical structure of 2-designs but emphasize
different aspects of physical dynamics: RB measures how noise \emph{breaks}
Haar invariance, whereas XEB measures how efficiently a circuit \emph{approaches}
it.  
RB characterizes the \emph{loss of ideality} under noise, while XEB characterizes
the \emph{emergence of chaos} in the evolution of pseudo-random circuits.
Random circuits that converge rapidly to 2-design behavior correspond to fast
scramblers: they spread local information quickly, delocalizing it across the
entire Hilbert space.
The rate at which \(F_{\mathrm{XEB}}\) increases with circuit depth therefore
provides an experimental proxy for the system’s scrambling efficiency.

}

\section{Conclusions}\label{sec:conclusions}

Random matrix theory and the invariant geometry of the unitary group under the Haar measure provide a unified
framework for understanding both static and dynamical signatures of quantum
chaos and information scrambling.  
Universal features of entanglement and spectral statistics follow naturally from
the geometry of \(U(D)\) together with concentration-of-measure effects in large
Hilbert spaces, explaining why Haar-based predictions so often describe generic
many-body systems.
Haar-random states serve as benchmarks for maximal entanglement and complexity;
the induced Wishart ensemble, the Marchenko--Pastur law, and Page’s theorem
characterize the entanglement structure of their subsystems; and dynamical
diagnostics such as the spectral form factor and out-of-time-ordered correlators connect these static features to
late-time signatures of chaos.

Although exact Haar randomness is exponentially costly, realistic chaotic
dynamics and shallow random circuits rapidly generate states whose reduced
statistics closely approximate those of Haar-random states.  
This effective emergence of randomness—often formalized through unitary
\(t\)-designs—provides the bridge between the ideal mathematical constructions
discussed here and the behavior of physically realizable many-body systems.

The same concepts have found practical expression in quantum computing, where
approximate unitary designs underpin widely used benchmarking protocols such as
randomized benchmarking~\cite{emerson2005scalable,magesan2011rb,magesan2012rb,Wallman2014}
and cross-entropy benchmarking~\cite{boixo2018XEB,arute2019quantum}.  
The abstract frameworks of Haar randomness and information scrambling have thus
evolved into concrete tools for the quantitative assessment of quantum hardware.
Randomized benchmarking translates the symmetries of the Haar measure into
practical estimators of average gate fidelity, while cross-entropy benchmarking
uses the statistical fingerprints of Haar-random states to gauge how faithfully
a device explores its Hilbert space.

These lecture notes are meant as an entry point rather than an exhaustive
treatment.  
Many directions invite further exploration, including quantitative rates of
convergence to Haar randomness, geometric measures of complexity, and the role
of entanglement-spectrum universality in quantum simulation and device
characterization.  
Each of these topics builds directly on the ideas introduced here and provides a
natural continuation for readers interested in the broader landscape of quantum
chaos, information dynamics, and characterization of quantum processors.

\section*{Acknowledgments}

We thank Grzegorz Rajchel-Mieldzio{\'c} and Timothy Heighthman for useful comments.

\bibliographystyle{apsrev4-2}
\bibliography{references}

@article{Sekino2008,
	title        = {Fast Scramblers},
	author       = {Sekino, Y. and Susskind, L.},
	year         = 2008,
	journal      = {Journal of High Energy Physics},
	volume       = 2008,
	number       = 10,
	pages        = {065},
	doi          = {10.1088/1126-6708/2008/10/065}
}

@article{Aitken1949,
  title = {On the Wishart Distribution in Statistics},
  volume = {36},
  ISSN = {0006-3444},
  url = {http://dx.doi.org/10.2307/2332529},
  DOI = {10.2307/2332529},
  number = {1/2},
  journal = {Biometrika},
  publisher = {JSTOR},
  author = {Aitken,  A. C.},
  year = {1949},
  month = jun,
  pages = {59}
}

@article{Nielsen2006,
  title = {Quantum Computation as Geometry},
  volume = {311},
  ISSN = {1095-9203},
  url = {http://dx.doi.org/10.1126/science.1121541},
  DOI = {10.1126/science.1121541},
  number = {5764},
  journal = {Science},
  publisher = {American Association for the Advancement of Science (AAAS)},
  author = {Nielsen,  Michael A. and Dowling,  Mark R. and Gu,  Mile and Doherty,  Andrew C.},
  year = {2006},
  month = feb,
  pages = {1133–1135}
}

@article{Maldacena2016,
	title        = {A bound on chaos},
	author       = {Maldacena,  Juan and Shenker,  Stephen H. and Stanford,  Douglas},
	year         = 2016,
	month        = aug,
	journal      = {Journal of High Energy Physics},
	publisher    = {Springer Science and Business Media LLC},
	volume       = 2016,
	number       = 8,
	doi          = {10.1007/jhep08(2016)106},
	issn         = {1029-8479},
	url          = {http://dx.doi.org/10.1007/JHEP08(2016)106}
}

@article{DAlessio2016,
	title        = {From quantum chaos and eigenstate thermalization to statistical mechanics and thermodynamics},
	author       = {D’Alessio,  Luca and Kafri,  Yariv and Polkovnikov,  Anatoli and Rigol,  Marcos},
	year         = 2016,
	month        = may,
	journal      = {Advances in Physics},
	publisher    = {Informa UK Limited},
	volume       = 65,
	number       = 3,
	pages        = {239–362},
	doi          = {10.1080/00018732.2016.1198134},
	issn         = {1460-6976},
	url          = {http://dx.doi.org/10.1080/00018732.2016.1198134}
}

@article{Mele2024,
	title        = {Introduction to Haar Measure Tools in Quantum Information: A Beginner&amp;apos;s Tutorial},
	author       = {Mele,  Antonio Anna},
	year         = 2024,
	month        = may,
	journal      = {Quantum},
	publisher    = {Verein zur Forderung des Open Access Publizierens in den Quantenwissenschaften},
	volume       = 8,
	pages        = 1340,
	doi          = {10.22331/q-2024-05-08-1340},
	issn         = {2521-327X},
	url          = {http://dx.doi.org/10.22331/q-2024-05-08-1340}
}

@BOOK{mehta2004random,
  title     = "Random matrices",
  author    = "Mehta, Madan L",
  publisher = "Academic Press",
  series    = "Pure and Applied Mathematics",
  edition   =  3,
  month     =  may,
  year      =  2014,
}

@article{Forrester1993,
	title        = {Log-gases,  random matrices and the Fisher-Hartwig conjecture},
	author       = {Forrester,  P J},
	year         = 1993,
	month        = mar,
	journal      = {Journal of Physics A: Mathematical and General},
	publisher    = {IOP Publishing},
	volume       = 26,
	number       = 5,
	pages        = {1179–1191},
	doi          = {10.1088/0305-4470/26/5/035},
	issn         = {1361-6447},
	url          = {http://dx.doi.org/10.1088/0305-4470/26/5/035}
}

@book{haake2010quantum,
	title        = {Quantum Signatures of Chaos},
	author       = {Haake, Fritz},
	year         = 2010,
	publisher    = {Springer},
	address      = {Berlin},
	edition      = {3rd}
}

@article{zyczkowski2001induced,
	title        = {Induced measures in the space of mixed quantum states},
	author       = {Życzkowski, Karol and Sommers, Hans-Jürgen},
	year         = 2001,
	journal      = {Journal of Physics A: Mathematical and General},
	volume       = 34,
	number       = 35,
	pages        = 7111,
	doi          = {10.1088/0305-4470/34/35/335}
}

@article{nadal2010statistical,
  title = {Statistical Distribution of Quantum Entanglement for a Random Bipartite State},
  volume = {142},
  ISSN = {1572-9613},
  url = {http://dx.doi.org/10.1007/s10955-010-0108-4},
  DOI = {10.1007/s10955-010-0108-4},
  number = {2},
  journal = {Journal of Statistical Physics},
  publisher = {Springer Science and Business Media LLC},
  author = {Nadal,  Celine and Majumdar,  Satya N. and Vergassola,  Massimo},
  year = {2011},
  month = jan,
  pages = {403–438}
}

@ARTICLE{brandao2016local,
  title     = "Local random quantum circuits are approximate polynomial-designs",
  author    = "Brand{\~a}o, Fernando G S L and Harrow, Aram W and Horodecki,
               Micha{\l}",
  journal   = "Commun. Math. Phys.",
  publisher = "Springer Science and Business Media LLC",
  volume    =  346,
  number    =  2,
  pages     = "397--434",
  month     =  sep,
  year      =  2016,
  language  = "en"
}

@article{page1993average,
	title        = {Average entropy of a subsystem},
	author       = {Page, Don N.},
	year         = 1993,
	journal      = {Physical Review Letters},
	volume       = 71,
	number       = 9,
	pages        = 1291,
	doi          = {10.1103/PhysRevLett.71.1291}
}

@article{lloyd1988black,
	title        = {Black Holes, Demons and the Loss of Coherence},
	author       = {Lloyd, Seth},
	year         = 1988,
	journal      = {Ph.D. thesis, Rockefeller University}
}

@article{deutsch1991quantum,
	title        = {Quantum statistical mechanics in a closed system},
	author       = {Deutsch, J. M.},
	year         = 1991,
	journal      = {Physical Review A},
	volume       = 43,
	number       = 4,
	pages        = 2046,
	doi          = {10.1103/PhysRevA.43.2046}
}

@article{srednicki1994chaos,
	title        = {Chaos and quantum thermalization},
	author       = {Srednicki, Mark},
	year         = 1994,
	journal      = {Physical Review E},
	volume       = 50,
	number       = 2,
	pages        = 888,
	doi          = {10.1103/PhysRevE.50.888}
}

@article{popescu2006entanglement,
	title        = {Entanglement and the foundations of statistical mechanics},
	author       = {Popescu, Sandu and Short, Anthony J. and Winter, Andreas},
	year         = 2006,
	journal      = {Nature Physics},
	volume       = 2,
	number       = 11,
	pages        = {754--758},
	doi          = {10.1038/nphys444}
}

@article{goldstein2006canonical,
	title        = {Canonical typicality},
	author       = {Goldstein, Sheldon and Lebowitz, Joel L. and Tumulka, Roderich and Zanghì, Nino},
	year         = 2006,
	journal      = {Physical Review Letters},
	volume       = 96,
	number       = 5,
	pages        = {050403},
	doi          = {10.1103/PhysRevLett.96.050403}
}

@article{hayden2007black,
	title        = {Black holes as mirrors: quantum information in random subsystems},
	author       = {Hayden, Patrick and Preskill, John},
	year         = 2007,
	journal      = {Journal of High Energy Physics},
	volume       = 2007,
	number       = {09},
	pages        = 120,
	doi          = {10.1088/1126-6708/2007/09/120}
}

@article{nahum2017quantum,
  title = {Operator Spreading in Random Unitary Circuits},
  author = {Nahum, Adam and Vijay, Sagar and Haah, Jeongwan},
  journal = {Phys. Rev. X},
  volume = {8},
  issue = {2},
  pages = {021014},
  numpages = {30},
  year = {2018},
  month = {Apr},
  publisher = {American Physical Society},
  doi = {10.1103/PhysRevX.8.021014},
  url = {https://link.aps.org/doi/10.1103/PhysRevX.8.021014}
}

@article{ho2018chaos,
	title        = {Chaos and quantum information scrambling in spin chains},
	author       = {Ho, Wen Wei and Abanin, Dmitry A.},
	year         = 2018,
	journal      = {Physical Review B},
	publisher    = {APS},
	volume       = 97,
	number       = 12,
	pages        = 124302
}

@article{haferkamp2022linear,
	title        = {Linear growth of complexity in random circuits},
	author       = {Haferkamp, Jonas and Faist, Philippe and Kothakonda, Naveen and Eisert, Jens and Hangleiter, Dominik},
	year         = 2022,
	journal      = {Nature Physics},
	publisher    = {Nature Publishing Group},
	volume       = 18,
	number       = 5,
	pages        = {528--532}
}

@book{ledoux2001concentration,
	title        = {The concentration of measure phenomenon},
	author       = {Ledoux, Michel},
	year         = 2001,
	publisher    = {American Mathematical Society}
}

@article{lashkari2013towards,
  title = {Towards the fast scrambling conjecture},
  volume = {2013},
  ISSN = {1029-8479},
  url = {http://dx.doi.org/10.1007/JHEP04(2013)022},
  DOI = {10.1007/jhep04(2013)022},
  number = {4},
  journal = {Journal of High Energy Physics},
  publisher = {Springer Science and Business Media LLC},
  author = {Lashkari,  Nima and Stanford,  Douglas and Hastings,  Matthew and Osborne,  Tobias and Hayden,  Patrick},
  year = {2013},
  month = apr,
}

@article{wishart1928generalised,
	title        = {The generalised product moment distribution in samples from a normal multivariate population},
	author       = {Wishart, John},
	year         = 1928,
	journal      = {Biometrika},
	volume       = {20A},
	number       = {1/2},
	pages        = {32--52}
}

@article{zyczkowski2003hilbert,
	title        = {Hilbert–Schmidt volume of the set of mixed quantum states},
	author       = {Zyczkowski, Karol and Sommers, Hans-J{\"u}rgen},
	year         = 2003,
	journal      = {Journal of Physics A: Mathematical and General},
	volume       = 36,
	number       = 39,
	pages        = 10115
}

@article{bengtsson2017geometry,
	title        = {Geometry of Quantum States: An Introduction to Quantum Entanglement},
	author       = {Bengtsson, Ingemar and Zyczkowski, Karol},
	year         = 2017,
	journal      = {Cambridge University Press}
}

@article{marchenko1967distribution,
	title        = {Distribution of eigenvalues for some sets of random matrices},
	author       = {Marchenko, Vladimir A. and Pastur, Leonid A.},
	year         = 1967,
	journal      = {Mathematics of the USSR-Sbornik},
doi = {10.1070/SM1967v001n04ABEH001994},
url = {https://doi.org/10.1070/SM1967v001n04ABEH001994},
	volume       = 1,
	number       = 4,
	pages        = 457
}

@article{hosur2016chaos,
	title        = {Chaos in quantum channels},
	author       = {Hosur, Pavan and Qi, Xiao-Liang and Roberts, Daniel A. and Yoshida, Beni},
	year         = 2016,
	journal      = {Journal of High Energy Physics},
	volume       = 2016,
	number       = 2,
	pages        = 4
}

@article{nahum2018operator,
	title        = {Operator spreading in random unitary circuits},
	author       = {Nahum, Adam and von Keyserlingk, Curtis W. and Vijay, Sagar and Haah, Jeongwan},
	year         = 2018,
	journal      = {Physical Review X},
	volume       = 8,
	number       = 2,
	pages        = {021014}
}

@article{dankert2009exact,
	title        = {Exact and approximate unitary 2-designs and their application to fidelity estimation},
	author       = {Dankert, Christoph and Cleve, Richard and Emerson, Joseph and Livine, Etera},
	year         = 2009,
	journal      = {Physical Review A},
	volume       = 80,
	number       = 1,
	pages        = {012304}
}

@article{dyson1962threefold,
	title        = {The threefold way. Algebraic structure of symmetry groups and ensembles in quantum mechanics},
	author       = {Dyson, Freeman J.},
	year         = 1962,
	journal      = {Journal of Mathematical Physics},
	publisher    = {AIP Publishing},
	volume       = 3,
	number       = 6,
	pages        = {1199--1215}
}

@article{bohigas1984characterization,
	title        = {Characterization of chaotic quantum spectra and universality of level fluctuation laws},
	author       = {Bohigas, Oriol and Giannoni, Marie-Joya and Schmit, Charles},
	year         = 1984,
	journal      = {Physical Review Letters},
	publisher    = {APS},
	volume       = 52,
	number       = 1,
	pages        = {1--4}
}

@article{tracy1994level,
	title        = {Level-spacing distributions and the Airy kernel},
	author       = {Tracy, Craig A. and Widom, Harold},
	year         = 1994,
	journal      = {Communications in Mathematical Physics},
	publisher    = {Springer},
	volume       = 159,
	number       = 1,
	pages        = {151--174}
}

@article{tracy1996orthogonal,
	title        = {On orthogonal and symplectic matrix ensembles},
	author       = {Tracy, Craig A. and Widom, Harold},
	year         = 1996,
	journal      = {Communications in Mathematical Physics},
	publisher    = {Springer},
	volume       = 177,
	number       = 3,
	pages        = {727--754}
}

@article{maldacena2016bound,
	title        = {A bound on chaos},
	author       = {Maldacena, Juan and Shenker, Stephen H. and Stanford, Douglas},
	year         = 2016,
	journal      = {Journal of High Energy Physics},
	publisher    = {Springer},
	volume       = 2016,
	number       = 8,
	pages        = 106
}

@ARTICLE{Harrow2023-xd,
  title     = "Approximate unitary t-designs by short random quantum circuits
               using nearest-neighbor and long-range gates",
  author    = "Harrow, Aram W and Mehraban, Saeed",
  journal   = "Commun. Math. Phys.",
  publisher = "Springer Science and Business Media LLC",
  volume    =  401,
  number    =  2,
  pages     = "1531--1626",
  month     =  jul,
  year      =  2023,
  copyright = "https://creativecommons.org/licenses/by/4.0",
  language  = "en"
}

@article{hunterjones2019unitary,
	title        = {Unitary designs from statistical mechanics in random quantum circuits},
	author       = {Hunter-Jones, Nicholas},
	year         = 2019,
	journal      = {Quantum},
	volume       = 3,
	pages        = 201
}

@article{susskind2016computational,
	title        = {Computational complexity and black hole horizons},
	author       = {Susskind, Leonard},
	year         = 2016,
	journal      = {Fortschritte der Physik},
	publisher    = {Wiley-VCH},
	volume       = 64,
	number       = 1,
	pages        = {24--43}
}

@article{brown2018secondlaw,
	title        = {The second law of quantum complexity},
	author       = {Brown, Adam R. and Susskind, Leonard},
	year         = 2018,
	journal      = {Physical Review D},
	publisher    = {APS},
	volume       = 97,
	number       = 8,
	pages        = {086015}
}

@article{roberts2017chaoscomplexity,
	title        = {Chaos and complexity by design},
	author       = {Roberts, Daniel A. and Yoshida, Beni},
	year         = 2017,
	journal      = {Journal of High Energy Physics},
	publisher    = {Springer},
	volume       = 2017,
	number       = 4,
	pages        = 121
}

@article{hayden2006aspects,
  title = {Randomizing Quantum States: Constructions and Applications},
  volume = {250},
  ISSN = {1432-0916},
  url = {http://dx.doi.org/10.1007/s00220-004-1087-6},
  DOI = {10.1007/s00220-004-1087-6},
  number = {2},
  journal = {Communications in Mathematical Physics},
  publisher = {Springer Science and Business Media LLC},
  author = {Hayden,  Patrick and Leung,  Debbie and Shor,  Peter W. and Winter,  Andreas},
  year = {2004},
  month = jul,
  pages = {371–391}
}

@article{reimann2007typicality,
	title        = {Typicality for generalized microcanonical ensembles},
	author       = {Reimann, Peter},
	year         = 2007,
	journal      = {Physical Review Letters},
	publisher    = {American Physical Society},
	volume       = 99,
	number       = 16,
	pages        = 160404,
	doi          = {10.1103/PhysRevLett.99.160404}
}

@article{mezzadri2007generate,
	title        = {How to generate random matrices from the classical compact groups},
	author       = {Mezzadri, Francesco},
	year         = 2007,
	journal      = {Notices of the AMS},
	volume       = 54,
	number       = 5,
	pages        = {592--604}
}

@book{hall2015lie,
	title        = {Lie Groups, Lie Algebras, and Representations: An Elementary Introduction},
	author       = {Hall, Brian C.},
	year         = 2015,
	publisher    = {Springer},
	address      = {New York},
	doi          = {10.1007/978-3-319-13467-3},
	edition      = {2nd}
}

@book{helgason1978differential,
	title        = {Differential Geometry, Lie Groups, and Symmetric Spaces},
	author       = {Helgason, Sigurdur},
	year         = 1978,
	publisher    = {Academic Press},
	address      = {New York}
}

@book{lee2013smooth,
	title        = {Introduction to Smooth Manifolds},
	author       = {Lee, John M.},
	year         = 2013,
	publisher    = {Springer},
	address      = {New York},
	doi          = {10.1007/978-1-4419-9982-5},
	edition      = {2nd}
}

@Inbook{ashtekar1999geometrical,
author="Ashtekar, Abhay
and Schilling, Troy A.",
editor="Harvey, Alex",
title="Geometrical Formulation of Quantum Mechanics",
bookTitle="On Einstein's Path: Essays in Honor of Engelbert Schucking",
year="1999",
publisher="Springer New York",
address="New York, NY",
pages="23--65",
isbn="978-1-4612-1422-9",
doi="10.1007/978-1-4612-1422-9_3",
url="https://doi.org/10.1007/978-1-4612-1422-9_3"
}

@article{Chu1996,
  title = {Geometry of the quantum complex projective spaceCP q (N)},
  volume = {72},
  ISSN = {1431-5858},
  url = {http://dx.doi.org/10.1007/s002880050233},
  DOI = {10.1007/s002880050233},
  number = {1},
  journal = {Zeitschrift f\"{u}r Physik C: Particles and Fields},
  publisher = {Springer Science and Business Media LLC},
  author = {Chu,  Chong-Sun and Ho,  Pei-Ming and Zumino,  Bruno},
  year = {1996},
  month = mar,
  pages = {163–170}
}

@article{brody2001geometric,
	title        = {Geometric quantum mechanics},
	author       = {Brody, Dorje C. and Hughston, Lane P.},
	year         = 2001,
	journal      = {Journal of Geometry and Physics},
	volume       = 38,
	number       = 1,
	pages        = {19--53},
	doi          = {10.1016/S0393-0440(00)00052-8}
}

@article{nielsen2006geometric,
	title        = {A geometric approach to quantum circuit lower bounds},
	author       = {Nielsen, Michael A.},
	year         = 2006,
	journal      = {Quantum Information and Computation},
	volume       = 6,
	number       = 3,
	pages        = {213--262},
	url          = {https://arxiv.org/abs/quant-ph/0502070}
}

@book{Schwichtenberg2018,
	title        = {Physics from Symmetry},
	author       = {Schwichtenberg,  Jakob},
	year         = 2018,
	journal      = {Undergraduate Lecture Notes in Physics},
	publisher    = {Springer International Publishing},
	doi          = {10.1007/978-3-319-66631-0},
	isbn         = 9783319666310,
	issn         = {2192-4805},
	url          = {http://dx.doi.org/10.1007/978-3-319-66631-0}
}

@article{Deutsch2018ETHreview,
	title        = {Eigenstate thermalization hypothesis},
	author       = {Deutsch, Joshua M.},
	year         = 2018,
	journal      = {Reports on Progress in Physics},
	volume       = 81,
	number       = 8,
	pages        = {082001},
	doi          = {10.1088/1361-6633/aac9f1},
	eprint       = {1805.01616},
	archiveprefix = {arXiv},
	primaryclass = {cond-mat.stat-mech}
}

@article{Hosur2016ChaosChannels,
	title        = {Chaos in quantum channels},
	author       = {Hosur, Pawel and Qi, Xiao-Liang and Roberts, Daniel A. and Yoshida, Beni},
	year         = 2016,
	journal      = {Journal of High Energy Physics},
	volume       = 2016,
	number       = 2,
	pages        = 4,
	doi          = {10.1007/JHEP02(2016)004},
	eprint       = {1511.04021},
	archiveprefix = {arXiv},
	primaryclass = {hep-th}
}

@article{Schmidt2022OTOCreview,
	title        = {Out-of-time-order correlators and quantum chaos: a review},
	author       = {Schmidt, Flavio and others},
	year         = 2022,
	journal      = {Scholarpedia},
	volume       = 17,
	number       = 3,
	pages        = 55205
}

@book{Tao2012TopicsRMT,
	title        = {Topics in Random Matrix Theory},
	author       = {Tao, Terence},
	year         = 2012,
	publisher    = {American Mathematical Society},
	series       = {Graduate Studies in Mathematics},
	volume       = 132,
	doi          = {10.1090/gsm/132}
}

@article{Liu2018,
	title        = {Entanglement,  quantum randomness,  and complexity beyond scrambling},
	author       = {Liu,  Zi-Wen and Lloyd,  Seth and Zhu,  Elton and Zhu,  Huangjun},
	year         = 2018,
	month        = jul,
	journal      = {Journal of High Energy Physics},
	publisher    = {Springer Science and Business Media LLC},
	volume       = 2018,
	number       = 7,
	doi          = {10.1007/jhep07(2018)041},
	issn         = {1029-8479},
	url          = {http://dx.doi.org/10.1007/JHEP07(2018)041}
}

@article{MaldacenaStanford2016CommentsSYK,
	title        = {Comments on the Sachdev–Ye–Kitaev model},
	author       = {Maldacena, Juan and Stanford, Douglas},
	year         = 2016,
	journal      = {Physical Review D},
	volume       = 94,
	number       = 10,
	pages        = 106002,
	doi          = {10.1103/PhysRevD.94.106002},
	eprint       = {1604.07818},
	archiveprefix = {arXiv},
	primaryclass = {hep-th}
}

@article{SachdevYe1993GaplessSpinFluid,
  title = {Gapless spin-fluid ground state in a random quantum Heisenberg magnet},
  author = {Sachdev, Subir and Ye, Jinwu},
  journal = {Phys. Rev. Lett.},
  volume = {70},
  issue = {21},
  pages = {3339--3342},
  numpages = {0},
  year = {1993},
  month = {May},
  publisher = {American Physical Society},
  doi = {10.1103/PhysRevLett.70.3339},
  url = {https://link.aps.org/doi/10.1103/PhysRevLett.70.3339}
}

@article{Chapman2022,
  title = {Quantum computational complexity from quantum information to black holes and back},
  volume = {82},
  ISSN = {1434-6052},
  url = {http://dx.doi.org/10.1140/epjc/s10052-022-10037-1},
  DOI = {10.1140/epjc/s10052-022-10037-1},
  number = {2},
  journal = {The European Physical Journal C},
  publisher = {Springer Science and Business Media LLC},
  author = {Chapman,  Shira and Policastro,  Giuseppe},
  year = {2022},
  month = feb 
}

@article{emerson2005scalable,
	title        = {Scalable noise estimation with random unitary operators},
	author       = {Emerson, Joseph and Alicki, Robert and Zyczkowski, Karol},
	year         = 2005,
	journal      = {Journal of Optics B: Quantum and Semiclassical Optics},
	volume       = 7,
	number       = 10,
	pages        = {S347},
	doi          = {10.1088/1464-4266/7/10/021}
}

@article{magesan2011rb,
	title        = {Scalable and robust randomized benchmarking of quantum processes},
	author       = {Magesan, Easwar and Gambetta, Jay M. and Emerson, Joseph},
	year         = 2011,
	journal      = {Physical Review Letters},
	volume       = 106,
	number       = 18,
	pages        = 180504,
	doi          = {10.1103/PhysRevLett.106.180504}
}

@article{boixo2018XEB,
	title        = {Characterizing quantum supremacy in near-term devices},
	author       = {Boixo, Sergio and Smelyanskiy, Vadim N. and Kechedzhi, Kostyantyn and others},
	year         = 2018,
	journal      = {Nature Physics},
	volume       = 14,
	pages        = {595--600},
	doi          = {10.1038/s41567-018-0124-x}
}

@article{arute2019quantum,
	title        = {Quantum supremacy using a programmable superconducting processor},
	author       = {Arute, Frank and Wu, K. and Google AI Quantum and collaborators},
	year         = 2019,
	journal      = {Nature},
	volume       = 574,
	pages        = {505--510},
	doi          = {10.1038/s41586-019-1666-5}
}

@article{emerson2007symmetrized,
	title        = {Symmetrized characterization of noisy quantum processes},
	author       = {Emerson, Joseph and Silva, Marcus and Moussa, Ossama and Ryan, Colm and Laforest, Martin and Baugh, Jonathan and Cory, David G and Laflamme, Raymond},
	year         = 2007,
	journal      = {Science},
	volume       = 317,
	number       = 5846,
	pages        = {1893--1896},
	doi          = {10.1126/science.1145699}
}

@article{magesan2012rb,
	title        = {Characterizing quantum gates via randomized benchmarking},
	author       = {Magesan, Easwar and Gambetta, Jay M and Emerson, Joseph},
	year         = 2012,
	journal      = {Physical Review A},
	volume       = 85,
	number       = 4,
	pages        = {042311},
	doi          = {10.1103/PhysRevA.85.042311}
}

@article{Wallman2014,
doi = {10.1088/1367-2630/16/10/103032},
url = {https://doi.org/10.1088/1367-2630/16/10/103032},
year = {2014},
month = {oct},
publisher = {IOP Publishing},
volume = {16},
number = {10},
pages = {103032},
author = {Wallman, Joel J and Flammia, Steven T},
title = {Randomized benchmarking with confidence},
journal = {New Journal of Physics},
}

@article{porter1956fluctuations,
	title        = {Fluctuations of nuclear reaction widths},
	author       = {Porter, C. E. and Thomas, R. G.},
	year         = 1956,
	journal      = {Physical Review},
	volume       = 104,
	number       = 2,
	pages        = {483--491},
	doi          = {10.1103/PhysRev.104.483}
}

@article{Swingle2016MeasuringScrambling,
	title        = {Measuring the scrambling of quantum information},
	author       = {Brian Swingle and Gregory Bentsen and Monika Schleier-Smith and Patrick Hayden},
	year         = 2016,
	journal      = {Physical Review A},
	volume       = 94,
	number       = 4,
	pages        = {040302},
	doi          = {10.1103/PhysRevA.94.040302},
	archiveprefix = {arXiv},
	eprint       = {1602.06271},
	primaryclass = {quant-ph}
}

@article{Swingle2018UnscramblingOTOC,
	title        = {Unscrambling the physics of out-of-time-order correlators},
	author       = {Brian Swingle},
	year         = 2018,
	journal      = {Nature Physics},
	volume       = 14,
	number       = 10,
	pages        = {988--990},
	doi          = {10.1038/s41567-018-0295-5}
}

@article{XuSwingle2024ScramblingOTOC,
	title        = {Scrambling Dynamics and Out-of-Time-Ordered Correlators in Quantum Many-Body Systems},
	author       = {Shenglong Xu and Brian Swingle},
	year         = 2024,
	journal      = {PRX Quantum},
	volume       = 5,
	number       = 1,
	pages        = {010201},
	doi          = {10.1103/PRXQuantum.5.010201},
	archiveprefix = {arXiv},
	eprint       = {2202.07060},
	primaryclass = {cond-mat.str-el}
}

@article{ShenkerStanford2014,
	title        = {Black holes and the butterfly effect},
	author       = {Shenker, Stephen H. and Stanford, Douglas},
	year         = 2014,
	journal      = {Journal of High Energy Physics},
	volume       = 2014,
	number       = 3,
	pages        = 67,
	doi          = {10.1007/JHEP03(2014)067}
}

@misc{Swingle2018Boulder,
	title        = {Boulder lectures on quantum information scrambling},
	author       = {Swingle, Brian},
	year         = 2018,
	url           = {https://boulderschool.yale.edu/sites/default/files/files/qi_boulder.pdf}
}

@article{Bekenstein1973BHentropy,
  title = {Black Holes and Entropy},
  author = {Bekenstein, Jacob D.},
  journal = {Phys. Rev. D},
  volume = {7},
  issue = {8},
  pages = {2333--2346},
  numpages = {0},
  year = {1973},
  month = {Apr},
  publisher = {American Physical Society},
  doi = {10.1103/PhysRevD.7.2333},
  url = {https://link.aps.org/doi/10.1103/PhysRevD.7.2333}
}

@article{Hawking1975,
	title        = {Particle creation by black holes},
	author       = {Hawking, S. W.},
	year         = 1975,
	journal      = {Communications in Mathematical Physics},
	volume       = 43,
	number       = 3,
	pages        = {199--220},
	doi          = {10.1007/BF02345020}
}

@article{Danshita2017SYKcoldatoms,
	title        = {Creating a Sachdev-Ye-Kitaev model in ultracold gases: From two-dimensional lattices to one-dimensional chains},
	author       = {Danshita, Ippei and Hanada, Masanori and Tezuka, Masaki},
	year         = 2017,
	journal      = {Progress of Theoretical and Experimental Physics},
	volume       = 2017,
	number       = 8,
	pages        = {083I01},
	doi          = {10.1093/ptep/ptx108}
}

@article{Wei2021,
  title = {Optical lattice platform for the Sachdev-Ye-Kitaev model},
  author = {Wei, Chenan and Sedrakyan, Tigran A.},
  journal = {Phys. Rev. A},
  volume = {103},
  issue = {1},
  pages = {013323},
  numpages = {9},
  year = {2021},
  month = {Jan},
  publisher = {American Physical Society},
  doi = {10.1103/PhysRevA.103.013323},
  url = {https://link.aps.org/doi/10.1103/PhysRevA.103.013323}
}

@article{Landsman2019IonOTOC,
	title        = {Verified Quantum Information Scrambling},
	author       = {Landsman, Kevin A. and others},
	year         = 2019,
	journal      = {Nature},
	volume       = 567,
	pages        = {61--65},
	doi          = {10.1038/s41586-019-0952-6}
}

@article{Chew2017GrapheneSYK,
	title        = {Approximating the Sachdev–Ye–Kitaev model with Majorana bilinears in graphene flakes},
	author       = {Chew, Aaron and Essin, Andrew and Alicea, Jason},
	year         = 2017,
	journal      = {Phys. Rev. B},
	volume       = 96,
	pages        = 121119,
	doi          = {10.1103/PhysRevB.96.121119}
}

@article{chan2018spectral,
	title        = {Spectral form factors and late-time quantum chaos},
	author       = {Chan, Amos M. and De Luca, Andrea and Chalker, J. T.},
	year         = 2018,
	journal      = {Phys. Rev. X},
	volume       = 8,
	pages        = {041019},
	doi          = {10.1103/PhysRevX.8.041019}
}

@article{zhang2019scrambling,
	title        = {Information Scrambling in Quantum Circuits},
	author       = {Zhang, Pengfei and Khemani, Vedika and Huse, David A.},
	year         = 2019,
	journal      = {Phys. Rev. B},
	volume       = 99,
	pages        = {075130},
	doi          = {10.1103/PhysRevB.99.075130}
}

@article{Bertini2020,
  title = {Scrambling in random unitary circuits: Exact results},
  author = {Bertini, Bruno and Piroli, Lorenzo},
  journal = {Phys. Rev. B},
  volume = {102},
  issue = {6},
  pages = {064305},
  numpages = {25},
  year = {2020},
  month = {Aug},
  publisher = {American Physical Society},
  doi = {10.1103/PhysRevB.102.064305},
  url = {https://link.aps.org/doi/10.1103/PhysRevB.102.064305}
}

@article{lieb1972finite,
	title        = {Finite group velocity of quantum interactions},
	author       = {Lieb, Elliott H. and Robinson, Derek},
	year         = 1972,
	journal      = {Commun. Math. Phys.}
}

@article{bravyi2006lieb,
  title = {Lieb-Robinson Bounds and the Generation of Correlations and Topological Quantum Order},
  volume = {97},
  ISSN = {1079-7114},
  url = {http://dx.doi.org/10.1103/PhysRevLett.97.050401},
  DOI = {10.1103/physrevlett.97.050401},
  number = {5},
  journal = {Physical Review Letters},
  publisher = {American Physical Society (APS)},
  author = {Bravyi,  S. and Hastings,  M. B. and Verstraete,  F.},
  year = {2006},
  month = jul 
}

@article{HaydenPreskill2007,
  title = {Black holes as mirrors: quantum information in random subsystems},
  volume = {2007},
  ISSN = {1029-8479},
  url = {http://dx.doi.org/10.1088/1126-6708/2007/09/120},
  DOI = {10.1088/1126-6708/2007/09/120},
  number = {09},
  journal = {Journal of High Energy Physics},
  publisher = {Springer Science and Business Media LLC},
  author = {Hayden,  Patrick and Preskill,  John},
  year = {2007},
  month = sep,
  pages = {120–120}
}

@article{DAlessioRigol2016,
	title        = {From quantum chaos and eigenstate thermalization to statistical mechanics},
	author       = {D'Alessio, Luca and Kafri, Yariv and Polkovnikov, Anatoli and Rigol, Marcos},
	year         = 2016,
	journal      = {Adv. Phys.},
	volume       = 65,
	pages        = {239--362}
}

@article{Abanin2019MBLreview,
  title = {Colloquium: Many-body localization, thermalization, and entanglement},
  author = {Abanin, Dmitry A. and Altman, Ehud and Bloch, Immanuel and Serbyn, Maksym},
  journal = {Rev. Mod. Phys.},
  volume = {91},
  issue = {2},
  pages = {021001},
  numpages = {26},
  year = {2019},
  month = {May},
  publisher = {American Physical Society},
  doi = {10.1103/RevModPhys.91.021001},
  url = {https://link.aps.org/doi/10.1103/RevModPhys.91.021001}
}

@article{Bertini2018,
  title = {Exact Spectral Form Factor in a Minimal Model of Many-Body Quantum Chaos},
  author = {Bertini, Bruno and Kos, Pavel and Prosen, Toma\ifmmode \check{z}\else \v{z}\fi{}},
  journal = {Phys. Rev. Lett.},
  volume = {121},
  issue = {26},
  pages = {264101},
  numpages = {6},
  year = {2018},
  month = {Dec},
  publisher = {American Physical Society},
  doi = {10.1103/PhysRevLett.121.264101},
  url = {https://link.aps.org/doi/10.1103/PhysRevLett.121.264101}
}

\appendix

{
 
\section{Mathematical structure and properties of the Haar measure}
\label{sec:haar_measure_detail}

The Haar measure provides the mathematical foundation for defining ``uniform randomness'' on continuous groups such as the unitary group $U(D)$.
For compact groups, such as $U(D)$, the Haar measure is both \emph{left-} and \emph{right-invariant} and can be normalized to unity.

Let $G$ be a compact topological group (for our purposes, $G = U(D)$, see App.\ref{app:topological_groups}).
A measure $d\mu(U)$ on $G$ is called a \emph{Haar measure} if it satisfies the invariance condition
\begin{equation}
d\mu(U) = d\mu(VU) = d\mu(UV), \qquad \forall\, V \in G.
\label{eq:haar_invariance}
\end{equation}
This invariance means that the measure assigns the same ``volume'' to any subset of $G$ and to its image under left or right multiplication by an arbitrary group element.
In other words, the Haar measure is completely uniform with respect to the group’s own symmetry structure.
It therefore represents the natural analogue of a uniform measure on finite groups or the Lebesgue measure on $\mathbb{R}^n$.

Because $U(D)$ is compact, the Haar measure can be normalized such that
\begin{equation}
\int_{U(D)} d\mu(U) = 1.
\label{eq:haar_normalization}
\end{equation}

Geometrically, $U(D)$ can be viewed as a manifold of all possible orthonormal bases in $\mathbb{C}^D$.
A unitary matrix $U$ corresponds to a rotation on this manifold, and the Haar measure defines a way to pick such a rotation uniformly at random.
Each infinitesimal ``volume element'' $d\mu(U)$ represents an equal share of this manifold’s total volume, ensuring that all directions in Hilbert space are equally probable.

The invariance property of Eq.~\eqref{eq:haar_invariance} implies that the Haar measure has no preferred basis or direction.
If one samples a unitary $U$ from $U(D)$ according to $d\mu(U)$, then both $U$ and $VU$ (for any fixed $V$) are distributed identically.
This invariance under both left and right multiplication is the mathematical expression of ``complete randomness'' in Hilbert space.

A Haar-random unitary $U$ acting on a fixed reference state $\ket{\psi_0}$ produces a Haar-random pure state:
\begin{equation}
\ket{\psi} = U \ket{\psi_0}.
\label{eq:haar_state}
\end{equation}
The resulting distribution of states is invariant under any unitary change of basis, which guarantees that e  direction on the unit sphere in $\mathbb{C}^D$ is equally likely.
Equivalently, one can describe the Haar measure on pure states as the uniform (unitarily invariant) measure on the complex unit sphere $S^{2D-1} \subset \mathbb{C}^D$:
\begin{equation}
d\mu(\psi) = \delta(1 - \braket{\psi}{\psi})\, d^{2D}\psi,
\label{eq:haar_state_measure}
\end{equation}
where $d^{2D}\psi$ denotes the flat measure on $\mathbb{C}^D$.

This construction allows random averages of quantum quantities to be defined rigorously, for example:
\begin{equation}
\mathbb{E}_U[f(U)] = \int_{U(D)} f(U)\, d\mu(U),
\end{equation}
or, equivalently for states,
\begin{equation}
\mathbb{E}_\psi[f(\psi)] = \int_{S^{2D-1}} f(\psi)\, d\mu(\psi).
\end{equation}

Exact sampling from the Haar measure can be achieved by generating a complex Ginibre random matrix $G$ with i.i.d.\ Gaussian entries and performing a QR decomposition~\cite{mezzadri2007generate}:
\begin{equation}
G = QR, \qquad U = Q D,
\end{equation}
where $D$ is a diagonal phase matrix ensuring $\det U = e^{i\theta}$.
The resulting $U$ is uniformly distributed according to the Haar measure on $U(D)$.
For large systems, approximate Haar-random unitaries can be realized efficiently using random local circuits or unitary $t$-designs~\cite{brandao2016local,dankert2009exact,hunterjones2019unitary}.

The Haar measure provides the natural benchmark for maximal entanglement and complete information scrambling.
Subsystems of Haar-random pure states exhibit fixed-trace Wishart statistics, their spectra follow the Marchenko--Pastur law, and their entanglement entropy satisfies Page’s theorem.
In this sense, the Haar ensemble defines the universal ``infinite-temperature'' limit of quantum chaos, against which physical dynamics and random circuits can be compared.

}

{
\section{Topological Groups and the Foundation of the Haar Measure}
\label{app:topological_groups}

A \emph{topology} on a set $X$ specifies which subsets of $X$ are considered ``open.''  
Formally, a topology is a collection $\mathcal{T}$ of subsets of $X$ that satisfies three axioms:
\begin{enumerate}
    \item Both the empty set $\varnothing$ and the entire set $X$ belong to $\mathcal{T}$.
    \item Arbitrary unions of open sets are open.
    \item Finite intersections of open sets are open.
\end{enumerate}
The pair $(X, \mathcal{T})$ is then called a \emph{topological space}.
Intuitively, a topology provides a notion of \emph{continuity}, \emph{closeness}, and \emph{convergence} without requiring a metric.
In physical terms, it defines the geometric structure that allows one to move smoothly from one point to another and to integrate over regions of the space. For example, $\mathbb{R}^n$ endowed with the standard Euclidean topology defines open sets as neighborhoods of points under the Euclidean distance.
Similarly, the set of complex unit vectors in $\mathbb{C}^D$ forms a topological space---the unit sphere $S^{2D-1}$---with the topology inherited from $\mathbb{C}^D$.

A \emph{group} is an algebraic structure $(G, \cdot)$ consisting of a set of elements and an associative binary operation that has an identity and inverses.
A \emph{topological group} is a group that is also a topological space, with the property that its multiplication and inversion operations are continuous maps:
\begin{align}
    &\text{Multiplication:} & m: G \times G &\to G, \qquad (g,h) \mapsto gh, \\
    &\text{Inversion:} & i: G &\to G, \qquad g \mapsto g^{-1}.
\end{align}
Continuity of these maps ensures that small changes in $g$ or $h$ result in small changes in $gh$ or $g^{-1}$.
In this way, algebraic operations on the group are compatible with its geometric structure. A familiar example is the additive group of real numbers $(\mathbb{R}, +)$, where both addition and negation are continuous.
In the quantum mechanical context, the most relevant example is the \emph{unitary group}
\[
U(D) = \{\,U \in \mathbb{C}^{D\times D} : U^\dagger U = \mathbb{I}_D\,\},
\]
which consists of all unitary matrices acting on a $D$-dimensional Hilbert space.
Matrix multiplication and Hermitian conjugation are continuous operations, and the set of unitary matrices is bounded and closed.
Hence $U(D)$ is a compact topological group.

The notion of topology is what makes it possible to define an integration measure on a group.
For a measure to exist, one must be able to talk about open sets, limits, and continuity of functions---all of which require a topological structure.

The Haar measure is defined on \emph{locally compact topological groups}, meaning that a  point has a neighborhood that is compact (finite total ``volume'').
In this setting, one can rigorously define integration of continuous functions over the group:
\begin{equation}
\int_G f(g)\, d\mu(g),
\end{equation}
and ask for a measure $d\mu(g)$ that is invariant under group multiplication:
\begin{equation}
d\mu(g) = d\mu(hg) = d\mu(gh), \qquad \forall h \in G.
\label{eq:haar_invariance_appendix}
\end{equation}
This invariance expresses the physical idea that the group has no preferred point or direction---a uniform probability measure on a continuous manifold.
The topological structure of $U(D)$ ensures that left and right translations of sets are continuous transformations, allowing the Haar measure to assign the same volume to all translated subsets.

Compactness plays a crucial role: for compact topological groups, such as $U(D)$ or $SU(D)$, the Haar measure is finite and can be normalized to unity,
\begin{equation}
\int_{U(D)} d\mu(U) = 1.
\end{equation}
Compactness ensures that averages over the group are well-defined, which is essential when defining random unitary operators or random pure states distributed uniformly in Hilbert space.

In the context of quantum information, the unitary group $U(D)$ forms the natural configuration space for all possible basis changes or quantum evolutions.
The Haar measure provides the only way to sample unitaries in a manner that is invariant under both left and right multiplication.
This ensures that all pure states $\ket{\psi} = U\ket{\psi_0}$ are equally likely, independent of the choice of reference state or computational basis.

Topological invariance thus encodes the fundamental symmetry of Hilbert space: there is no privileged direction or origin.
The Haar measure exploits this symmetry to define a uniform ``volume'' element on $U(D)$, from which all ensemble averages of random states or operators follow.
This structure is indispensable for formulating quantum typicality, random matrix theory, and the statistical description of information scrambling.

The Haar measure therefore exists precisely because $U(D)$ is a compact topological group: its algebraic and geometric properties jointly guarantee the existence of a uniform, invariant notion of randomness in quantum mechanics.

\subsection*{The general linear group and non-compactness}

The general linear group
\[
\mathrm{GL}(D,\mathbb{C}) = \{\, A \in \mathbb{C}^{D\times D} \mid \det A \neq 0 \,\}
\]
consists of all invertible complex matrices.
It is a topological group under matrix multiplication and inversion, but it is \emph{not compact}.
Compactness requires the space to be both bounded and closed, while $\mathrm{GL}(D,\mathbb{C})$ fails both conditions:
\begin{itemize}
    \item Matrices can have arbitrarily large norm, e.g.\ $\mathrm{diag}(t,1,\ldots,1)$ with $t\!\to\!\infty$.
    \item Limits of invertible matrices may become singular, e.g.\ $\mathrm{diag}(1,\epsilon)\!\to\!\mathrm{diag}(1,0)$ as $\epsilon\!\to\!0$.
\end{itemize}
Consequently, the Haar measure on $\mathrm{GL}(D,\mathbb{C})$ exists but cannot be normalized because its total volume is infinite.
In contrast, the unitary group $U(D)$ is both closed and bounded---e  element satisfies $U^\dagger U = \mathbb{I}$---making it compact and allowing its Haar measure to be normalized to unity.
This compactness underlies the existence of a finite, uniform notion of randomness for quantum states and unitaries.

}

{
 
\section{Length of a path and metric invariance on \texorpdfstring{$U(D)$}{U(D)}}
\label{app:length_of_path}

Given the Riemannian metric
\begin{equation}
\langle X, Y \rangle = -\mathrm{Tr}(X Y),
\qquad X,Y \in \mathfrak{u}(D),
\label{eq:metric_trace}
\end{equation}
we can define the length of a smooth path \(U(t)\in U(D)\) as
\begin{equation}
L[U(t)] = \int_0^1 \sqrt{\langle \dot{U}U^\dagger, \dot{U}U^\dagger \rangle}\, dt.
\label{eq:length_def}
\end{equation}
The factor \(\dot{U}U^\dagger\) represents the tangent vector in the Lie algebra \(\mathfrak{u}(D)\) obtained by left-translating the velocity \(\dot{U}\) back to the identity.

Differentiating the unitarity condition \(U^\dagger U = \mathbb{I}\) yields
\begin{equation}
\dot{U}^\dagger U + U^\dagger \dot{U} = 0,
\quad \Rightarrow \quad
U^\dagger \dot{U} \in \mathfrak{u}(D),
\label{eq:unitarity_derivative}
\end{equation}
so \(U^\dagger \dot{U}\) is skew-Hermitian.
Because the metric \eqref{eq:metric_trace} is invariant under left translations,
\[
\langle U X U^\dagger, U Y U^\dagger \rangle = \langle X, Y \rangle,
\]
we can compute the inner product at the identity:
\[
\langle \dot{U}U^\dagger, \dot{U}U^\dagger \rangle
= -\mathrm{Tr}\!\big[(\dot{U}U^\dagger)(\dot{U}U^\dagger)\big].
\]
Since \((\dot{U}U^\dagger)^\dagger = -(\dot{U}U^\dagger)\), we have
\[
-\mathrm{Tr}\!\big[(\dot{U}U^\dagger)^2\big]
= \mathrm{Tr}\!\big[(\dot{U}U^\dagger)^\dagger (\dot{U}U^\dagger)\big]
= \mathrm{Tr}(\dot{U}^\dagger \dot{U}),
\]
where the last equality follows from cyclicity of the trace.

Substituting this into Eq.~\eqref{eq:length_def} gives the compact and manifestly invariant expression
\begin{equation}
L[U(t)] = \int_0^1 \sqrt{\mathrm{Tr}\!\big(\dot{U}^\dagger \dot{U}\big)}\, dt.
\label{eq:length_final}
\end{equation}

The integrand in Eq.~\eqref{eq:length_final} plays the role of a ``speed'' on the manifold:
\[
v(t) = \sqrt{\mathrm{Tr}(\dot{U}^\dagger \dot{U})}.
\]
It measures how fast the system moves through the space of unitaries.
For a time-independent Hamiltonian \(H\), with \(U(t) = e^{-iHt}\), one finds
\[
\dot{U} = -iHU,
\qquad
\mathrm{Tr}(\dot{U}^\dagger \dot{U}) = \mathrm{Tr}(H^2),
\]
so the motion occurs along a geodesic of constant speed, and the total length grows linearly with time.
This identifies Schrödinger evolution as a uniform rotation on the compact manifold \(U(D)\).

The left-invariant metric \eqref{eq:metric_trace} thus endows \(U(D)\) with a natural notion of distance and curvature, allowing one to treat quantum dynamics geometrically as motion on a compact Riemannian manifold of constant positive curvature.

}

{
\section{Riemannian Metric on \texorpdfstring{$SU(2)$}{SU(2)}}
\label{app:su2_metric}

In this appendix we explicitly derive the Riemannian line element on \(SU(2)\) 
using the invariant metric introduced in Eq.~\eqref{eq:riemannian_metric}.
This computation illustrates how the general formalism of the unitary group \(U(D)\)
reduces, in the simplest nontrivial case, to the geometry of the three-sphere \(S^3\).

Every element of \(SU(2)\) can be parameterized by four real angles \((\alpha,\beta,\gamma,\delta)\) as
\begin{equation}
U(\alpha,\beta,\gamma,\delta) 
= e^{i\alpha}
\begin{pmatrix}
e^{i\beta}\cos\gamma & e^{i\delta}\sin\gamma \\
-\,e^{-i\delta}\sin\gamma & e^{-i\beta}\cos\gamma
\end{pmatrix}.
\label{eq:U2_param}
\end{equation}
The global phase \(e^{i\alpha}\) corresponds to the \(U(1)\) subgroup and does not affect
distances on \(SU(2)\); we therefore omit it in what follows and define
\[
V(\beta,\gamma,\delta)
=
\begin{pmatrix}
e^{i\beta}\cos\gamma & e^{i\delta}\sin\gamma \\
-\,e^{-i\delta}\sin\gamma & e^{-i\beta}\cos\gamma
\end{pmatrix}.
\]

\vspace{0.3em}
\noindent
The infinitesimal variation of \(V\) under small changes of the parameters is
\[
dV =
\frac{\partial V}{\partial \beta}\, d\beta
+\frac{\partial V}{\partial \gamma}\, d\gamma
+\frac{\partial V}{\partial \delta}\, d\delta,
\]
with partial derivatives
\begin{align}
\frac{\partial V}{\partial \beta}
&= i
\begin{pmatrix}
e^{i\beta}\cos\gamma & 0\\[2pt]
0 & -e^{-i\beta}\cos\gamma
\end{pmatrix},\\[3pt]
\frac{\partial V}{\partial \gamma}
&=
\begin{pmatrix}
-\,e^{i\beta}\sin\gamma & e^{i\delta}\cos\gamma\\[2pt]
-\,e^{-i\delta}\cos\gamma & -\,e^{-i\beta}\sin\gamma
\end{pmatrix},\\[3pt]
\frac{\partial V}{\partial \delta}
&= i
\begin{pmatrix}
0 & e^{i\delta}\sin\gamma\\[2pt]
e^{-i\delta}\sin\gamma & 0
\end{pmatrix}.
\end{align}

\vspace{0.3em}
\noindent
The bi-invariant Riemannian metric on \(U(D)\) is
\[
ds^2 = \mathrm{Tr}(dU^{\dagger} dU),
\]
which for the present case becomes \(ds^2 = \mathrm{Tr}(dV^{\dagger} dV)\).
Expanding \(dV\) in the differential basis \((d\beta,d\gamma,d\delta)\) yields
\[
ds^2 = \sum_{i,j} 
\mathrm{Tr}\!\left(
\frac{\partial V^{\dagger}}{\partial x_i}
\frac{\partial V}{\partial x_j}
\right) 
dx_i\,dx_j,
\qquad x_i,x_j \in \{\beta,\gamma,\delta\}.
\]

The metric coefficients are defined as
\[
g_{ij} = \mathrm{Tr}\!\left(
\frac{\partial V^{\dagger}}{\partial x_i}
\frac{\partial V}{\partial x_j}
\right).
\]

Each off-diagonal element is zero, i.e. for \(i\neq j\), all traces vanish:
\[
g_{ij} = 0 \quad \text{for } i \neq j,
\]
because  each derivative 
\(\partial_\beta V\), \(\partial_\gamma V\), and \(\partial_\delta V\)
corresponds to motion along a distinct tangent direction on the manifold,
generated by different elements of the Lie algebra \(\mathfrak{su}(2)\),
which is spanned by the orthogonal Pauli matrices.
Under the Hilbert--Schmidt inner product
\(\langle A,B\rangle = \mathrm{Tr}(A^{\dagger}B)\),
the Pauli generators satisfy \(\mathrm{Tr}(\sigma_i\sigma_j)=2\delta_{ij}\),
implying that tangent directions associated with different parameters are orthogonal.
This ensures that all cross terms vanish, so the metric is diagonal in these coordinates. As such, metric contains only diagonal elements. Straightforward computation gives
\begin{align}
g_{\gamma\gamma} &= \mathrm{Tr}\!\left(
\frac{\partial V^{\dagger}}{\partial \gamma}
\frac{\partial V}{\partial \gamma}
\right)
= 2, \\[3pt]
g_{\beta\beta} &= \mathrm{Tr}\!\left(
\frac{\partial V^{\dagger}}{\partial \beta}
\frac{\partial V}{\partial \beta}
\right)
= 2\cos^2\gamma, \\[3pt]
g_{\delta\delta} &= \mathrm{Tr}\!\left(
\frac{\partial V^{\dagger}}{\partial \delta}
\frac{\partial V}{\partial \delta}
\right)
= 2\sin^2\gamma.
\end{align}
After normalizing by an overall factor of two,
the infinitesimal line element becomes
\begin{equation}
ds^2 = d\gamma^2 + \cos^2\gamma\, d\beta^2 + \sin^2\gamma\, d\delta^2.
\label{eq:su2_metric}
\end{equation}

Equation~\eqref{eq:su2_metric} coincides with the standard round metric on the three-sphere \(S^3\) of unit radius. 
Each element of \(SU(2)\) thus corresponds to a point on \(S^3\), and the coordinates
\((\beta,\gamma,\delta)\) serve as generalized spherical coordinates.
The orthogonality of the coordinate directions reflects the fact that the tangent vectors
generated by the Pauli matrices are mutually perpendicular in the Lie-algebra sense,
so geodesic motion in one parameter direction does not mix with the others.
This confirms the geometric equivalence \(SU(2) \simeq S^3\),
and provides an explicit demonstration of how the Hilbert--Schmidt metric on \(U(D)\)
reduces to the familiar geometry of a three-sphere in the simplest case.

}

\section{Distribution of coefficients of Haar-random States}
\label{app:haar_gaussian}

Let $\ket{\psi}$ be a pure state chosen uniformly at random according to the Haar measure on the unit sphere in $\mathbb{C}^{mn}$,
\[
\ket{\psi} = \sum_{i=1}^{m}\sum_{\mu=1}^{n} C_{i\mu}\,\ket{i}_A \otimes \ket{\mu}_B,
\qquad
\sum_{i,\mu}|C_{i\mu}|^2 = 1.
\]
Then the joint probability density of the coefficients $\{C_{i\mu}\}$ is invariant under all unitary transformations of the basis, and the marginal distribution of any finite subset of coefficients coincides with that of independent complex Gaussian variables with zero mean and variance $1/(mn)$.

The Haar measure on the unit sphere $S^{2mn-1} \subset \mathbb{C}^{mn}$ is the unique unitarily invariant probability measure.
This invariance implies that for any unitary matrix $U \in U(mn)$,
\[
\ket{\psi} \mapsto U\ket{\psi}
\]
leaves the measure unchanged.  
Consequently, the probability density $P(\{C_{i\mu}\})$ must depend only on the rotationally invariant quantity
\(\sum_{i,\mu} |C_{i\mu}|^2.\)

The most general such distribution is
\begin{equation}
P(\{C_{i\mu}\}) \propto
\delta\!\left(1 - \sum_{i,\mu}|C_{i\mu}|^2\right),
\label{eq:haar_sphere}
\end{equation}
which defines a uniform measure on the hypersphere of radius one in $\mathbb{C}^{mn}$.
Equation~\eqref{eq:haar_sphere} can equivalently be realized by drawing each $C_{i\mu}$ as an independent complex Gaussian random variable with variance $\sigma^2$ and then normalizing the resulting vector:
\[
C_{i\mu} = \frac{Z_{i\mu}}{\sqrt{\sum_{j,\nu}|Z_{j\nu}|^2}},
\quad
Z_{i\mu} \sim \mathcal{N}_{\mathbb{C}}(0,\sigma^2),
\]
where $\mathcal{N}_{\mathbb{C}}(0,\sigma^2)$ denotes a complex Gaussian with zero mean and variance $\sigma^2$ per complex component.
The denominator enforces the normalization $\sum_{i,\mu}|C_{i\mu}|^2 = 1$.

Since the normalization factor is independent of direction in $\mathbb{C}^{mn}$, this construction reproduces the uniform measure on the sphere.
Choosing $\sigma^2 = 1/(mn)$ ensures that $\mathbb{E}[|C_{i\mu}|^2] = 1/(mn)$ after normalization.

Therefore, the coefficients of a Haar-random pure state can be treated as independent complex Gaussian variables with zero mean and variance $1/(mn)$, up to the global normalization constraint.
In matrix notation, this implies that the coefficient matrix $C$ is distributed as a complex Ginibre random matrix~\cite{Forrester1993,nadal2010statistical,hayden2006aspects} constrained by $\mathrm{Tr}(C C^{\dagger}) = 1$.

{
\section{Random Matrix Theory and spectral statistics}
\label{app:RMT_intro}

Random matrix theory (RMT) is a simple but powerful framework for understanding
typical spectral features of complicated quantum systems.  
The basic idea dates back to Wigner, who proposed that when a Hamiltonian is so
complicated that no analytic solution is possible—as in heavy nuclei—one can
replace it by a large matrix with random entries, provided the random matrix
respects the same global symmetries as the original system.  
Surprisingly, this crude approximation turns out to reproduce very accurately
the universal patterns observed in chaotic spectra
\cite{haake2010quantum}.

A natural place to begin is with the distinction between integrable and chaotic
systems.  An integrable quantum system possesses many conserved quantities, so
its energy levels do not strongly influence one another.  
Empirically, the spacings between consecutive levels behave much like
independent random numbers: for ordered energies
\(E_1 \le E_2 \le \cdots\), the spacings
\(s_n = E_{n+1}-E_n\) follow the exponential (Poisson) distribution
\[
  P_{\mathrm{Poi}}(s) = \frac{1}{\bar{s}} e^{-s/\bar{s}},
\]
where \(\bar{s}\) is the mean spacing.  
A key feature is the finite value \(P_{\mathrm{Poi}}(0)\): levels do not repel
each other, so small spacings are common.

Chaotic systems behave differently.  
Their energy levels avoid coincidences, a phenomenon known as \emph{level
repulsion}.  After removing the smooth background variation of the level
density—an operation called \emph{unfolding}—the spacings follow the
Wigner--Dyson distributions of the three classical Gaussian random matrix
ensembles:
\[
P(s) \propto s^\beta e^{-c s^{2}} ,
\]
with \(\beta=1,2,4\) and \(c\) a constant.  
The power-law factor \(s^\beta\) encodes level repulsion: the probability of
finding two nearly degenerate levels vanishes as a power of their separation.
These three values of \(\beta\) correspond to three symmetry classes, to which
almost every Hermitian Hamiltonian belongs.

The parameter \(\beta\) arises from the way a Hamiltonian transforms under
time-reversal.  
A time-reversal operator \(T\) is an \emph{antiunitary} map (i.e.\ it performs
complex conjugation, possibly followed by a unitary transformation).  
If \(THT^{-1}=H\), the system is time-reversal invariant.  
Two possibilities occur: \(T^2=+1\), typical for spinless or integer-spin
particles, and \(T^2=-1\), which leads to Kramers degeneracy in half-integer
spin systems.  
Hamiltonians with \(T^2=+1\) can be represented by real-symmetric matrices and
belong to the \emph{Gaussian Orthogonal Ensemble} (GOE, \(\beta=1\));
Hamiltonians with no time-reversal symmetry belong to the
\emph{Gaussian Unitary Ensemble} (GUE, \(\beta=2\));
and Hamiltonians with \(T^2=-1\) fall into the
\emph{Gaussian Symplectic Ensemble} (GSE, \(\beta=4\))
\cite{dyson1962threefold}.
The threefold way provides a simple and robust classification: to understand
typical spectral behaviour, one needs only the symmetry class, not the detailed
form of the Hamiltonian.

One of the central mathematical results of RMT is that the repulsion exponent
\(\beta\) has a geometric origin.  
For many invariant probability measures on Hermitian matrices, the transition
from matrix entries to eigenvalues introduces a Jacobian factor
\[
  \prod_{i<j} |\lambda_i - \lambda_j|^{\beta},
\]
where \(\lambda_i\) denote the eigenvalues.  
Thus the eigenvalues behave like particles that repel each other with a
logarithmic interaction.  
This viewpoint is made precise in the ``Coulomb gas'' analogy: the joint density
of eigenvalues can be written as
\begin{equation}
P(\lambda_1,\ldots,\lambda_N)
  \propto
  \exp\!\left[
       -\beta\left(
          - \sum_{i<j} \log|\lambda_i-\lambda_j|
          + \frac{1}{2}\sum_i V(\lambda_i)
       \right)
    \right],
    \label{eq:P_joint}
\end{equation}
where \(V(\lambda)\) is an effective confining potential.  
The logarithmic term enforces level repulsion, while the potential determines
the overall shape of the spectrum.  
Importantly, many fine details of \(V\) do not matter for local statistics:
the repulsion law, the spacing distribution, and the correlation functions are
\emph{universal} within each symmetry class.

The parameter $\beta$ appearing in the joint eigenvalue distribution, Eq.\eqref{eq:P_joint},
is known as the \emph{Dyson index},
underlies the ``threefold way'' classification of invariant ensembles Ref.~\cite{mehta2004random,Forrester1993,haake2010quantum,bohigas1984characterization,zyczkowski2001induced,nadal2010statistical}.
It specifies the symmetry class of the random matrix ensemble and controls the strength of eigenvalue correlations.  
The Dyson index counts the number of real degrees of freedom per independent matrix element.

There exist three canonical symmetry classes corresponding to the possible values of $\beta$:
\begin{equation}
\beta =
\begin{cases}
1, & \text{Real-symmetric (Orthogonal ensemble, GOE)},\\[4pt]
2, & \text{Complex-Hermitian (Unitary ensemble, GUE)},\\[4pt]
4, & \text{Quaternion self-dual (Symplectic ensemble, GSE)}.
\end{cases}
\end{equation}

When transforming from matrix elements to eigenvalues and eigenvectors, the Jacobian of this change of variables generates the Vandermonde factor
\begin{equation}
J(\{\lambda_i\}) = \prod_{i<j} |\lambda_i - \lambda_j|^{\beta},
\end{equation}
which encodes the curvature of the space of Hermitian matrices and introduces correlations between eigenvalues.
The exponent $\beta$ thus determines the degree of \emph{level repulsion}:
\begin{equation}
P(s) \propto s^{\beta}, \qquad (s \to 0),
\end{equation}
where $P(s)$ is the nearest-neighbor level spacing distribution.
For $\beta=1$ (orthogonal), the distribution grows linearly near $s=0$; for $\beta=2$ (unitary), it grows quadratically; and for $\beta=4$ (symplectic), quartically.
In contrast, uncorrelated (Poissonian) spectra exhibit $P(s)=e^{-s}$ without level repulsion.

Within the Coulomb-gas formulation, the joint distribution can be written as
\begin{equation}
\begin{split}
P(\{\lambda_i\}) &\propto e^{-\beta {H}(\{\lambda_i\})},\\
{H} &= -\sum_{i<j}\log|\lambda_i - \lambda_j| + \frac{1}{2}\sum_i V(\lambda_i),
\end{split}
\end{equation}
where ${H}$ acts as the Hamiltonian of a one-dimensional gas of logarithmically repelling particles.
In this analogy, $\beta^{-1}$ plays the role of an \emph{effective temperature}:  
larger $\beta$ corresponds to a ``colder'' gas with stronger correlations and reduced fluctuations of eigenvalues.
Spectral rigidity therefore increases with $\beta$.

For the fixed-trace complex Wishart ensemble associated with reduced density matrices of Haar-random states, the matrix $X$ has complex entries and belongs to the unitary class, corresponding to $\beta=2$.  
Consequently, the entanglement spectrum of a random subsystem exhibits the same universal level statistics as the Gaussian Unitary Ensemble (GUE).
The quadratic level repulsion, $P(s)\propto s^2$ at small spacings, signifies strong eigenvalue correlations---a universal feature of chaotic quantum systems and fully scrambled quantum states.

}
 
\section{Derivation of the Joint Eigenvalue Distribution}
\label{app:eig_dist}

In this appendix we outline the derivation of the joint probability density for the eigenvalues of the reduced density matrix $\rho_A$ of a Haar-random pure state, following Refs.~\cite{zyczkowski2001induced,nadal2010statistical,Forrester1993,haake2010quantum}. As shown in Sec.~\ref{sec:rdm_haar}, the reduced density matrix $\rho_A$ can be written in the form
\begin{equation}
\rho_A = \frac{W}{\Tr W}, \qquad
W = X X^{\dagger},
\end{equation}
where $X$ is an $m\times n$ complex matrix with i.i.d.\ Gaussian entries of zero mean and unit variance.
The random matrix $W$ thus belongs to the complex Wishart ensemble with probability density
\begin{equation}
P(W)\,dW \propto (\det W)^{n - m} e^{-\Tr W}\, dW,
\label{eq:app_wishart}
\end{equation}
where $dW$ denotes the flat measure on the space of $m\times m$ Hermitian matrices.

Any Hermitian matrix $W$ can be diagonalized as
\begin{equation}
W = U \Lambda U^{\dagger},
\qquad
\Lambda = \mathrm{diag}(\lambda_1,\ldots,\lambda_m),
\end{equation}
where $U \in U(m)$ is unitary and $\lambda_i \ge 0$ are the eigenvalues.
The differential $dW$ transforms as
\begin{equation}
dW = J(\{\lambda_i\})\, d\mu(U) \prod_{i=1}^{m} d\lambda_i,
\end{equation}
where $d\mu(U)$ is the normalized Haar measure on $U(m)$ and
$J(\{\lambda_i\})$ is the Jacobian associated with the transformation.

For complex Hermitian matrices (the $\beta=2$ case of the Dyson index),
the Jacobian is given by the square of the Vandermonde determinant:
\begin{equation}
J(\{\lambda_i\}) = \prod_{i<j}|\lambda_i - \lambda_j|^2.
\label{eq:app_vandermonde}
\end{equation}
This factor captures the curvature of the space of Hermitian matrices and leads to eigenvalue repulsion.

Since the Wishart probability density~\eqref{eq:app_wishart} depends only on the invariants $\Tr W$ and $\det W$, the integral over the unitary group
contributes only a constant normalization factor.
Integrating over $U$ yields the joint probability density for the eigenvalues of $W$:
\begin{equation}
P(\lambda_1,\ldots,\lambda_m)
\propto
\prod_{i=1}^{m} \lambda_i^{n - m} e^{-\lambda_i}
\prod_{i<j} (\lambda_i - \lambda_j)^2.
\label{eq:app_wishart_eig}
\end{equation}
Equation~\eqref{eq:app_wishart_eig} is the standard result for the eigenvalue distribution of the complex Wishart ensemble.

The reduced density matrix $\rho_A = W / \Tr W$ must satisfy the normalization condition $\Tr \rho_A = 1$.
To impose this constraint, we insert a Dirac delta function into Eq.~\eqref{eq:app_wishart_eig}:
\begin{equation}
P(\lambda_1,\ldots,\lambda_m)
\propto
\delta\!\left(\sum_{i=1}^{m}\lambda_i - 1\right)
\prod_{i=1}^{m}\lambda_i^{n - m}
\prod_{i<j}(\lambda_i - \lambda_j)^2.
\label{eq:app_fixedtrace}
\end{equation}
Equation~\eqref{eq:app_fixedtrace} defines the \emph{fixed-trace Wishart ensemble}, also known as the \emph{induced ensemble} of density matrices~\cite{zyczkowski2001induced,bengtsson2017geometry}.
  This expression provides the exact joint distribution of eigenvalues of $\rho_A$ for Haar-random pure states~\cite{zyczkowski2001induced,bengtsson2017geometry,nadal2010statistical}. The delta function enforces unit trace, the factors $\lambda_i^{n - m}$ encode the relative subsystem–environment dimensions, and the Vandermonde term
$\prod_{i<j}(\lambda_i - \lambda_j)^2$ generates eigenvalue repulsion characteristic of unitary-invariant random matrix ensembles.
This structure underlies the universality of entanglement spectra and enables the analytical computation of ensemble-averaged quantities such as purity, Rényi entropies, and the von Neumann entropy.

To understand the physical content of Eq.~\eqref{eq:app_fixedtrace}, it is convenient to take its logarithm:
\begin{align}
\log P
= \text{const}
+ (n-m)\sum_{i}\log\lambda_i
+ 2\sum_{i<j}\log|\lambda_i - \lambda_j|.
\end{align}
Interpreting $-\log P$ as an energy functional defines an effective Hamiltonian
for ``particles'' located at positions $\{\lambda_i\}$ on the positive real axis:
\begin{equation}
{H}(\{\lambda_i\})
= - (n-m)\sum_i \log \lambda_i
- 2\sum_{i<j} \log|\lambda_i - \lambda_j|.
\label{eq:coulomb_energy}
\end{equation}
The first term represents a confining potential that keeps the eigenvalues within the unit interval due to normalization, while the second term describes pairwise logarithmic repulsion between them.  
This mapping identifies the eigenvalues as a classical one-dimensional \emph{Coulomb gas} or ``log-gas''~\cite{Forrester1993,haake2010quantum}.  
Each eigenvalue behaves as a charged particle interacting with all others via a repulsive potential
\[
U_{ij} = -2\log|\lambda_i - \lambda_j|.
\]
Because the Coulomb potential in one dimension is logarithmic, the interaction is long-ranged and enforces strong correlations among the eigenvalues. The factor $\prod_{i<j}(\lambda_i - \lambda_j)^2$ ensures that configurations with coinciding eigenvalues have zero measure:  
\[
P(\lambda_i=\lambda_j) = 0.
\]
This \emph{level repulsion} suppresses small spacings and gives rise to
\emph{spectral rigidity}---the tendency of eigenvalues to avoid clustering.
For small spacings $s = |\lambda_{i+1}-\lambda_i|$, the nearest-neighbor
spacing distribution behaves as
\[
P(s) \propto s^{\beta}, \qquad \beta = 2,
\]
for the complex (unitary) ensemble. In contrast, an uncorrelated (Poisson)
spectrum has
\[
P_{\mathrm{Pois}}(s) = e^{-s},
\]
which does not vanish at $s=0$. Thus $P(s\!\to\!0)\to 0$ is a characteristic
signature of Wigner--Dyson ensembles and reflects the logarithmic repulsion
between eigenvalues.

On larger scales, rigidity is quantified by the \emph{number variance}
$\Sigma^{2}(L)$, defined as
\[
\Sigma^{2}(L) = \big\langle (N(L) - L)^2 \big\rangle.
\]
Here the spectrum has been \emph{unfolded}, meaning that each eigenvalue
$\lambda_i$ is mapped to a new variable
\[
x_i = \bar N(\lambda_i),
\]
where $\bar N(\lambda)$ is the smooth integrated density of states
(the ensemble-averaged eigenvalue counting function).
This transformation removes slow variations in the average density,
so that the unfolded levels $\{x_i\}$ have unit mean spacing.  The quantity
$N(L)$ then denotes the number of unfolded levels in an interval of length~$L$.
For Wigner--Dyson spectra, $\Sigma^{2}(L)$ grows sublinearly with $L$, whereas
for Poisson spectra one has $\Sigma^{2}(L)=L$. This reduced growth expresses
the rigidity of correlated random-matrix spectra.

In the context of spectral properties of reduced density matrices of Haar-random pure state, the the Coulomb-gas picture implies that the eigenvalues of $\rho_A$---the entanglement spectrum---cannot cluster arbitrarily.  
The normalization constraint $\sum_i \lambda_i = 1$ acts as an overall charge-neutrality condition, while the logarithmic repulsion distributes the eigenvalues evenly within their allowed domain.  
This mechanism explains why the entanglement spectrum of a random subsystem is ``smooth'' and universal, following the Marchenko--Pastur law in the large-$n$ limit.

Analogous phenomena occur in the energy-level statistics of chaotic quantum Hamiltonians, where eigenvalue correlations give rise to the Wigner--Dyson distribution~\cite{haake2010quantum,bohigas1984characterization}.

\section{Concentration of Measure and Typicality}\label{sec:typicality}

In high-dimensional Hilbert spaces, the overwhelming majority of pure states are locally indistinguishable.
This geometric concentration explains why Haar-random states and chaotic eigenstates share thermal properties at the subsystem level.
The term \emph{concentration of measure} refers to the geometric fact that on high-dimensional spaces---such as the unit sphere of a large Hilbert space---the uniform (Haar) measure assigns almost all its weight to points whose properties lie exponentially close to the average.
Consequently, smooth functions on such spaces take nearly the same value for almost all points, which explains why most pure states exhibit identical local properties.

The universality of entanglement and spectral statistics in random states originates from the geometry of high-dimensional Hilbert spaces.
Most points on a high-dimensional unit sphere are nearly orthogonal and exhibit similar local properties, a phenomenon known as the \emph{concentration of measure}~\cite{ledoux2001concentration,hayden2006aspects,popescu2006entanglement}.
In essence, as the Hilbert-space dimension $D$ grows, almost e  pure state is ``typical'' in the sense that observables restricted to small subsystems take nearly identical values for almost all states.

{
An essential geometric fact underlying the typicality of random pure states is that in high-dimensional Hilbert spaces, almost all normalized vectors are nearly orthogonal to each other.
This property---a manifestation of the concentration of measure---is the foundation for the statistical uniformity of subsystems in random quantum states. 
The probability distribution of the squared overlap 
\(\chi = |\inner{\psi}{\phi}|^2\) between two independent Haar-random pure states in $\mathbb{C}^D$
can be obtained directly from the geometry of the complex unit sphere $S^{2D-1}$.
Because the Haar measure is unitarily invariant, one may fix one of the states, say
\(\ket{\psi} = (1,0,\ldots,0)\), without loss of generality.  
Then for a random state $\ket{\phi}$ uniformly distributed on the unit sphere,
\begin{equation}
\chi = |\inner{\psi}{\phi}|^2 = |\phi_1|^2,
\label{eq:chi_component}
\end{equation}
where $\phi_1$ denotes the first complex component of $\ket{\phi}$.

The vector $\ket{\phi} = (\phi_1, \phi_2, \ldots, \phi_D)$ represents a normalized state in $\mathbb{C}^D$,
so its components satisfy the normalization condition
\[
\sum_{i=1}^{D} |\phi_i|^2 = 1.
\] 
They can therefore be regarded as coordinates inside a $(D-1)$-dimensional simplex—the space of all possible sets of nonnegative numbers that add up to unity.

Within this simplex, the joint probability density of the components $\{|\phi_i|^2\}$ is flat; all configurations consistent with the normalization are equally likely.
This uniform distribution is mathematically known as the \emph{Dirichlet distribution} with all parameters equal to one:
\begin{equation}
P(|\phi_1|^2,\ldots,|\phi_D|^2)
\propto
\delta\!\left(1 - \sum_{i=1}^{D}|\phi_i|^2\right).
\label{eq:dirichlet_simplex}
\end{equation}

Integrating out all but the first coordinate yields the marginal distribution for
$\chi = |\phi_1|^2$:
\begin{equation}
P(\chi)
\propto
(1 - \chi)^{D-2}, \qquad 0 \le \chi \le 1.
\label{eq:chi_unscaled}
\end{equation}
The normalization condition $\int_0^1 P(\chi)\,d\chi = 1$ fixes the prefactor,
giving the exact result
\begin{equation}
P(\chi) = (D-1)(1 - \chi)^{D-2}, \qquad 0 \le \chi \le 1.
\label{eq:overlap_distribution_final},
\end{equation}
which is a Beta distribution,
\(\chi \sim \mathrm{Beta}(1, D-1)\).
The mean overlap is therefore
\begin{equation}
\mathbb{E}[\chi] = \int_0^1 \chi P(\chi)\, d\chi = \frac{1}{D}.
\label{eq:mean_overlap},
\end{equation}
and variance
\begin{equation}
\mathrm{Var}(\chi) = \frac{1}{D^2(D+1)}.
\label{eq:chi_moments}
\end{equation}
Thus, for large $D$, two randomly chosen normalized states are almost orthogonal on average, with typical overlap magnitude
\begin{equation}
| \inner{\psi}{\phi} | \sim \mathcal{O}(D^{-1/2}).
\label{eq:overlap_scaling}
\end{equation}

Equation~\eqref{eq:overlap_distribution_final} shows that the overlaps between random pure states are sharply peaked near $\chi=0$ and become increasingly concentrated around zero as the Hilbert-space dimension $D$ grows.
This behavior underlies the statement that almost all states in high-dimensional Hilbert space are \emph{nearly orthogonal}, i.e the probability that the overlap exceeds a small threshold $\epsilon$ decays exponentially \cite{ledoux2001concentration},
\begin{equation}
\Pr(|\inner{\psi|\phi}|^2 > \epsilon) \le e^{-c D \epsilon},
\label{eq:overlap_concentration}
\end{equation}
where $c$ is a positive constant of order unity.
The set of normalized vectors on the unit sphere in  $C^D$ 
 is so vast that nearly all of them are almost orthogonal to one another.
That means any direction in Hilbert space is typical — there are no privileged directions.

Because almost all pure states in high-dimensional Hilbert space are mutually nearly orthogonal,
they sample the unit sphere almost uniformly.  
When restricted to a small subsystem, the partial trace over the vast environment averages out their global differences,
so the corresponding reduced density matrices $\rho_A^{(\psi)}$ and $\rho_A^{(\phi)}$
become exponentially close to each other.  
Hence, distinct global pure states yield almost identical local properties—a manifestation of canonical typicality. The fact that almost all pure states in high-dimensional Hilbert space are mutually nearly orthogonal has a profound physical implication: it guarantees that their local properties are almost indistinguishable when only a small subsystem is observed.

Consider two independent Haar-random pure states $\ket{\psi}$ and $\ket{\phi}$ in a bipartite Hilbert space $\mathcal{H} = \mathcal{H}_A \otimes \mathcal{H}_B$ with dimensions $\dim \mathcal{H}_A = m$ and $\dim \mathcal{H}_B = n$, so that the total dimension is $D = mn$.
Their overlap is typically exponentially small,
\begin{equation}
|\braket{\psi}{\phi}|^2 \sim \frac{1}{D}.
\label{eq:smalloverlap}
\end{equation}
Nevertheless, the reduced density matrices
\[
\rho_A^{(\psi)} = \mathrm{Tr}_B\!\left[\ket{\psi}\bra{\psi}\right],
\qquad
\rho_A^{(\phi)} = \mathrm{Tr}_B\!\left[\ket{\phi}\bra{\phi}\right],
\]
are almost identical when $n \gg m$.
This follows because the partial trace acts as a strong contraction: differences that are large in the full Hilbert space are washed out after tracing over the environment.
More formally, for any observable $\mathcal{O}_A$ acting nontrivially only on subsystem $A$, the expectation value
\begin{equation}
X = \bra{\psi}\mathcal{O}_A\ket{\psi}
\end{equation}
is sharply concentrated around its mean value
\begin{equation}
\mathbb{E}[X] = \mathrm{Tr}\!\left(\mathcal{O}_A \frac{\mathbb{I}_A}{m}\right),
\label{eq:meanvalue}
\end{equation}
which corresponds to the maximally mixed state on $\mathcal{H}_A$.
By Lévy’s lemma~\cite{ledoux2001concentration,hayden2006aspects,popescu2006entanglement}, the probability of a deviation larger than $\epsilon$ satisfies
\begin{equation}
\Pr(|X - \mathbb{E}[X]| > \epsilon)
\le
2\exp[-C D \epsilon^2 / \|\mathcal{O}_A\|^2],
\label{eq:levybound}
\end{equation}
where $C$ is an $\mathcal{O}(1)$ constant and $\|\mathcal{O}_A\|$ the operator norm.
This exponential suppression with the total Hilbert-space dimension $D$ implies that for almost e  random pure state,
\begin{equation}
\rho_A^{(\psi)} \approx \frac{\mathbb{I}_A}{m},
\label{eq:localmixed}
\end{equation}
up to exponentially small corrections. 
Equation~\eqref{eq:localmixed} states that the reduced state of any small subsystem drawn from a Haar-random pure state is nearly maximally mixed, independent of the specific global state.
Hence, although different global states are almost perfectly distinguishable (orthogonal) when viewed in the full Hilbert space, they are \emph{locally indistinguishable} when restricted to small subsystems.
This is the essence of \emph{canonical typicality}~\cite{popescu2006entanglement,hayden2006aspects}:
the local properties of almost all pure states coincide with those of the thermal state $\mathbb{I}_A/m$.

The near-orthogonality of global states and the near-identity of local states are thus two sides of the same geometric phenomenon.
Both stem from the concentration of measure in high-dimensional spaces, which ensures that almost e  pure state exhibits the same local observables and entanglement properties.
This observation provides the geometric foundation of the eigenstate thermalization hypothesis~\cite{deutsch1991quantum,srednicki1994chaos}, explaining why individual eigenstates of chaotic Hamiltonians appear thermal when viewed locally.

This concentration property also underpins the Eigenstate Thermalization Hypothesis (ETH)~\cite{deutsch1991quantum,srednicki1994chaos}.
Even for individual energy eigenstates of chaotic Hamiltonians, the local reduced density matrices closely approximate the microcanonical ensemble.
In this sense, Page’s theorem and the ETH are two complementary manifestations of the same geometric phenomenon: the overwhelming uniformity of high-dimensional quantum state space.

}

The concentration of measure phenomenon formalizes the intuition that in   high-dimensional spaces, almost all points are close to each other with respect to any ``smooth'' function.
A central result capturing this idea is \emph{Lévy’s lemma}~\cite{ledoux2001concentration,hayden2006aspects,popescu2006entanglement}.
Let $f : S^N \to \mathbb{R}$ be a Lipschitz-continuous function on the $N$-dimensional unit sphere $S^N$ in $\mathbb{R}^{N+1}$.
That is, there exists a constant $\eta_f$ such that for all $x,y \in S^N$,
\begin{equation}
|f(x) - f(y)| \le \eta_f \, \|x - y\|.
\label{eq:lipschitz}
\end{equation}
If $x$ is drawn uniformly at random from the sphere (with respect to the Haar measure), then for any $\epsilon > 0$,
\begin{equation}
\Pr\!\left(|f(x) - \mathbb{E}[f]| > \epsilon\right)
\le
2 \exp\!\left(-\frac{c N \epsilon^2}{\eta_f^2}\right),
\label{eq:levy_math}
\end{equation}
where $c$ is a positive constant of order unity.
This inequality states that the probability that $f(x)$ deviates from its mean $\mathbb{E}[f]$ by more than $\epsilon$ is exponentially small in the dimension $N$. Equation~\eqref{eq:levy_math} encapsulates the idea that as $N$ grows,
any smooth function of many independent variables becomes almost constant:
fluctuations vanish exponentially with dimension.

In the context of quantum mechanics, we consider normalized pure states $\ket{\psi}$ in a Hilbert space $\mathbb{C}^D$.
The set of all such states forms the complex unit sphere $S^{2D-1}$ equipped with the Haar measure.
For any observable $\mathcal{O}_A$ acting on a subsystem $A$, define the function
\begin{equation}
f(\psi) = \bra{\psi}\mathcal{O}_A\ket{\psi}.
\label{eq:fpsi}
\end{equation}
The function $f(\psi)$ is Lipschitz with constant $\eta_f = 2\|\mathcal{O}_A\|$, since
\[
|f(\psi) - f(\phi)| \le 2\|\mathcal{O}_A\| \, \|\psi - \phi\|.
\]
Applying Lévy’s lemma with $N = 2D - 1$ yields
\begin{equation}
\Pr\!\left(|\bra{\psi}\mathcal{O}_A\ket{\psi} - \mathbb{E}[f]| > \epsilon\right)
\le
2\exp\!\left[-C D \epsilon^2 / \|\mathcal{O}_A\|^2\right],
\label{eq:levy_quantum}
\end{equation}
where $C$ is an $\mathcal{O}(1)$ constant.
The mean value $\mathbb{E}[f]$ corresponds to the expectation over the maximally mixed state on the subsystem:
\begin{equation}
\mathbb{E}[f] = \mathrm{Tr}\!\left(\mathcal{O}_A \frac{\mathbb{I}_A}{m}\right).
\label{eq:levy_mean}
\end{equation}

Equation~\eqref{eq:levy_quantum} shows that the expectation value of any local observable $\mathcal{O}_A$ is sharply concentrated around its average for almost all pure states.
Deviations of order $\epsilon$ occur with exponentially small probability $\sim \exp(-c D \epsilon^2)$. Consequently, for every pure state $\ket{\psi}$,
the reduced density matrix of subsystem $A$ satisfies
\begin{equation}
\rho_A^{(\psi)} \approx \frac{\mathbb{I}_A}{m},
\label{eq:levy_local}
\end{equation}
up to corrections exponentially suppressed in the total Hilbert-space dimension $D$.
This is the precise statement of \emph{canonical typicality}:  
most pure states are locally indistinguishable from the maximally mixed state. Almost all pure states of large quantum systems exhibit locally thermal behavior, 
providing a rigorous framework for understanding why small subsystems of high-dimensional quantum states appear maximally mixed---a geometric consequence of measure concentration on the Hilbert-sphere
\cite{ledoux2001concentration,popescu2006entanglement,hayden2006aspects,goldstein2006canonical,reimann2007typicality}.

Implication of Lévy’s lemma is the fact that typicality   property is purely geometric—it requires no assumptions about dynamics or equilibrium.
The apparent thermality of small subsystems is a direct consequence of the high-dimensional geometry of quantum state space.
It provides the mathematical foundation for both Page’s theorem and the Eigenstate Thermalization Hypothesis, which assert that individual chaotic eigenstates are locally thermal.

{
The concentration of measure phenomenon can be understood most transparently through geometry.
To illustrate the concept, we begin with the unit sphere in three dimensions and then generalize to high-dimensional Hilbert spaces.
Let us consider the ordinary unit sphere $S^2 \subset \mathbb{R}^3$ and define a ``height'' function along the $z$-axis,
\begin{equation}
f(x,y,z) = z.
\end{equation}
A point at the north pole has $z=1$, one at the south pole has $z=-1$, and the equator corresponds to $z=0$.
If points are sampled uniformly from the sphere, the probability density of $z$ is uniform between $-1$ and $1$:
\begin{equation}
P(z) = \frac{1}{2}, \qquad -1 \le z \le 1.
\end{equation}
Thus, on an ordinary sphere, the surface area is evenly distributed between the poles and the equator---there is no ``concentration'' yet.

Now, let us consider the unit sphere in $N$ dimensions, $S^{N-1} \subset \mathbb{R}^N$, and again choose one coordinate axis, say $x_1$.
For a uniformly random point $\mathbf{x} = (x_1,\ldots,x_N)$ on the sphere, the distribution of $x_1$ is~\cite{ledoux2001concentration}
\begin{equation}
P(x_1) \propto (1 - x_1^2)^{(N-3)/2}, \qquad -1 \le x_1 \le 1.
\label{eq:highdim_x1}
\end{equation}
As the dimension $N$ increases, the factor $(1 - x_1^2)^{(N-3)/2}$ strongly suppresses large $|x_1|$, producing a sharp peak around $x_1 = 0$.
The variance of $x_1$ is
\begin{equation}
\mathrm{Var}(x_1) = \frac{1}{N},
\end{equation}
so the typical value of $|x_1|$ scales as $1/\sqrt{N}$.
Although $x_1$ can in principle take values between $-1$ and $1$, for $N = 1000$ it almost always lies within $|x_1| \lesssim 0.03$.

Geometrically all points on the high-dimensional sphere lie within 
a   thin \textit{equatorial} band
around the hyperplane $x_1 = 0$, perpendicular to the chosen axis.
The surface area of this equatorial region grows exponentially with $N$, while the area near the poles shrinks exponentially.
As a result, the uniform (Haar) measure on the sphere assigns almost all of its weight to this thin equatorial region. 
{Independently on how the equator is rotated, leading to a great circle, almost all points lie in its vicinity. On multidimensional Earth-like planet, that would correspond to the fact that almost all points have tropical climate, but also that almost all of them lie in the same timezone.}

The same geometry applies to the complex unit sphere of pure quantum states $S^{2D-1} \subset \mathbb{C}^D$.
Each normalized state $\ket{\psi}$ corresponds to a point on this sphere.
A local observable $\mathcal{O}_A$ defines a smooth function
\[
f(\psi) = \bra{\psi}\mathcal{O}_A\ket{\psi}.
\]
Lévy’s lemma ensures that for large $D$, this function takes nearly the same value for almost all $\ket{\psi}$.
The overwhelming majority of quantum states therefore lie in the ``equatorial band'' of the Hilbert sphere---that is, the region where expectation values of local observables are exponentially close to their average.

The equatorial band picture explains why almost all pure states of large quantum systems appear locally identical.
Although global states are nearly orthogonal to one another, their projections onto small subsystems fall within this narrow region of similar local statistics.
Hence, reduced density matrices $\rho_A$ are nearly indistinguishable from the maximally mixed state $\mathbb{I}_A/m$, and local observables take almost identical expectation values.
This geometric concentration of measure is therefore the underlying reason why Haar-random states, chaotic eigenstates, and typical many-body states all exhibit the same local thermal properties.
}

\end{document}